\documentclass{aa}
\usepackage{graphicx} 
\usepackage[varg]{txfonts}
\usepackage{orcidlink}
\usepackage{placeins}
\usepackage{hyperref}
\hypersetup{
    colorlinks=true,
    linkcolor=blue,
    filecolor=magenta,      
    urlcolor=blue,
    citecolor=blue,
    pdftitle={Overleaf Example},
    }
\bibpunct{(}{)}{;}{a}{}{,} 
\usepackage{booktabs}
\usepackage{float}
\usepackage{subcaption}
\usepackage{placeins}
\usepackage[labelfont=bf, font=footnotesize]{caption}

\begin{document}

\title{Incidence of afterglow plateaus in gamma-ray bursts associated with binary neutron star mergers}

\author{L. Guglielmi\inst{\ref{i1},\ref{i2}}\orcidlink{0009-0008-4255-8033}\and G. Stratta\inst{\ref{i3},\ref{i4},\ref{i5},\ref{i6}}\orcidlink{0000-0003-1055-7980}\and S. Dall'Osso\inst{\ref{i6},\ref{i7}}\orcidlink{0000-0003-4366-8265}\and P. Singh\inst{\ref{i3},\ref{i5}}\orcidlink{0000-0003-1006-6970}\and M. Brusa\inst{\ref{i2},\ref{i5}}\orcidlink{0000-0002-5059-6848}\and R. Perna\inst{\ref{i8}}\orcidlink{0000-0002-3635-5677} }

\institute{
Department of Aerospace Engineering, Politecnico di Torino, Corso Duca degli Abruzzi 24, Torino, Italia, 10129 \label{i1} 
\and
Dipartimento di Fisica e Astronomia “Augusto Righi”, Università di Bologna, via Gobetti 93/2, 40129 Bologna, Italy \label{i2} 
\and
Institut für Theoretische Physik, Goethe Universität, Max-von-Laue-Str. 1, D-60438 Frankfurt am Main, Germany \label{i3} 
\and
Istituto di Astrofisica e Planetologia Spaziali, via Fosso del Cavaliere 100, I-00133 Roma, Italy \label{i4} 
\and
INAF—Osservatorio Astronomico di Bologna, viale P.Gobetti, Bologna, Italy \label{i5} 
\and
Istituto Nazionale di Fisica Nucleare—Roma 1, Piazzale Aldo Moro 2, I-00185 Roma, Italy \label{i6} \and
``Sapienza" Universit\`{a} di Roma, Dipartimento di Fisica, Piazzale Aldo Moro 2, I-00185 Roma, Italy \label{i7}
\and
Department of Physics \& Astronomy, Stony Brook University, Stony Brook, NY 11794-3800, US \label{i8}
}

\date{Received 13 August 2024 /
       Accepted 3 October 2024}

\abstract{One of the most surprising gamma-ray burst (GRB) features discovered with the \textit{Swift} X-ray telescope (XRT) is a plateau phase in 
the early X-ray afterglow light curves.  
These plateaus are observed in the majority of long GRBs,  
while their incidence in short GRBs (SGRBs) 
is still uncertain due to their fainter X-ray afterglow luminosity with respect to long GRBs. An accurate estimate of the fraction of SGRBs with plateaus  
is of utmost relevance given the implications that the plateau may have for our understanding of the jet structure 
and possibly of the nature of the binary neutron star (BNS) merger remnant.
This work presents the results of an extensive data analysis of the largest and most up-to-date sample of SGRBs observed with the XRT, and for which the redshift has been measured. We find a plateau incidence of 18-37\% in SGRBs, which is a significantly lower fraction than that measured in long GRBs (>50\%). 
Although still 
debated, the plateau phase 
could be explained as energy injection from the spin-down power of a newly born magnetized neutron star (NS; magnetar).
We show that 
this scenario can nicely reproduce 
the observed short GRB (SGRBs) plateaus, while at the same time providing a natural explanation for the different plateau fractions between short and long GRBs. In particular, our findings may imply that only a minority of BNS mergers generating SGRBs leave behind a sufficiently stable or long-lived NS  to form a plateau. 
From the 
probability distribution
of the BNS remnant mass, a fraction 18-37\% of short GRB plateaus implies a maximum NS mass in the range $\sim 2.3-2.35$ M$_{\odot}$.
}


\maketitle

\section{Introduction}
\label{sec:intro}

Gamma-ray bursts (GRBs) have been a great astrophysical mystery since their discovery in the late 1960s. By the end of the 1990s, their cosmological origin was assessed with the discovery of the 
afterglow component (e.g. \citealt{costa1997}) and the identification of their host galaxies \citep{Metzger1997}.
A new breakthrough came with the launch of the Neil Gehrels Swift Observatory (\textit{Swift} hereafter, \citealt{gehrelsetal2004}) in November 2004, which allowed the first observations of the early phases (a few minutes after the burst) of the afterglow emission, leading to the discovery of unexpected features that are thought to encode crucial information on the jet structure and possibly on the nature of the remnant compact object. 

More specifically, the  early observations of the \textit{Swift}/X-ray telescope (XRT) revealed, in most cases, an initial steep flux decay, likely marking the switching off of the prompt emission, followed by a shallow phase (the so-called plateau), which then transitions to a characteristic power-law flux decay \citep{ZhangB_2006ApJ...642..354Z}. While the latter is in agreement with the afterglow theory of synchrotron emission by electrons energised in a relativistic shock \citep{sari1998}, the plateau could not be explained in the same framework, requiring additional physics.

After almost two decades of \textit{Swift}/XRT GRB observations, we now know that plateaus occur in the majority of long GRBs, which are those associated with the collapse of massive stars. Short GRBs (SGRBs) associated with binary neutron star (BNS) mergers (and possibly neutron star-black hole mergers), proved harder to study due to their fainter afterglow luminosity with respect to long GRBs: to date, the frequency of plateaus in SGRB afterglows is uncertain \citep{rowlinson2013}. 
A precise estimate of plateau incidence in SGRBs is of utmost relevance due to its potential impact on our understanding of their jet morphology 
and conceivably of the nature of the BNS merger remnant. Indeed, it has been suggested that plateaus could originate from geometrical effects in structured jets\footnote{See also  \citet{Dereli2022NatCo..13.5611D} for an alternative interpretation (for a small sample of long GRBs) based on a low bulk Lorentz factor and a low-density wind medium.} (e.g. \citealt{Oganesyan_2020, Beniamini_2022}). 
In this case,  the plateau incidence in short and long GRBs is expected to 
be comparable, 
since geometrical effects are of a similar nature in both types.
An alternative interpretation invokes the formation of a neutron star (NS) remnant injecting energy into the forward shock \citep{usov1992,Dai2,Dai1,Fan1,Metzger_2007, dallosso2011,Ronchini1}. A fascinating consequence of this scenario is that the incidence of 
SGRB plateaus potentially reflects 
the fraction of BNS systems that form a NS remnant after the merger. 
The latter, in turn, can provide information on the maximum mass of a stable NS (see e.g. \citealt{Sarin2,Sarin1}).  
 

This work is divided into two parts: first, we examine the X-ray afterglow light-curve morphology of the most complete sample of SGRBs at known redshift to date, adopting a completely agnostic approach and aiming to confidently identify or rule out the presence of a plateau in each analysed SGRB. We then place our results in a more physical context, assuming that the plateaus are produced by magnetar remnants.

The sample of SGRBs is presented in Sect.~\ref{sec:sample} and the data analysis in Sect.~\ref{sec:analysis}, while the interpretation within the magnetar framework is elaborated in Sect.~\ref{sec:magnetar}. Some astrophysical implications are derived in Sect.~\ref{sec:maxmasstop}, with discussion and conclusions following in  Sects.~\ref{sec:discussion} and \ref{sec:conclusion}, respectively.

\section{The sample}
\label{sec:sample}

\begin{figure}
\resizebox{\hsize}{!}{\includegraphics{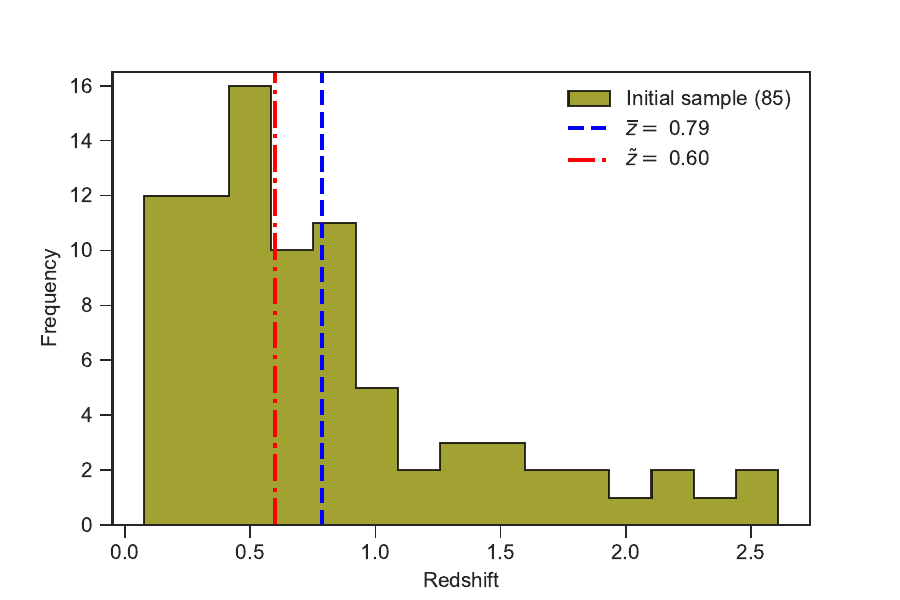}}
\caption{Redshift distribution of our sample of 85 SGRBs, spanning the range $z=[0.0763 - 2.609]$, with average $\overline{z}=0.79$ and median $\tilde{z}= 0.60$.
}
\label{fig:zhistini}
\end{figure}

The sample analysed in this work is composed of 85 SGRBs and includes all the SGRBs at known redshift detected by \textit{Swift} from May 2005 to the end of December 2021 (about 60\% of the total \textit{Swift} SGRB population; \citealt{fong2022}). The sample was built based on past works \citep{rossi2020, fong2022, oconnor2022}, where potential contamination from SGRBs associated with collapsars (e.g. GRB 200826, \citealt{Rossi_2022}) and long GRBs associated with BNSs was
taken into account. 
The list of SGRBs we obtained using observations with the X-ray
telescope XRT on board \textit{Swift} represents the most up to date and complete sample of SGRBs at known redshift presently available.  Figure~\ref{fig:zhistini} shows the redshift distribution of the sample, spanning the redshift range $z=[0.0763 - 2.609]$, with a median of $\langle z\rangle= 0.60$ and an average of $\overline{z}=0.79$.


\section{Data analysis}
\label{sec:analysis}

For each SGRB in the sample, the XRT 0.3-10 keV unabsorbed afterglow flux light curve was retrieved from the publicly available 
\textit{Swift} XRT Repository on the UK \textit{Swift} Science Data Center website \citep{evans2007,evans2009}. 
For visualisation purposes only, we  
also considered the \textit{Swift} Burst Alert Telescope (BAT), with 15-150 keV unabsorbed flux, 
to obtained a more detailed and complete  light curve from the prompt to the afterglow emission. 
We then analysed each XRT light curve as explained below. 

\subsection{Sample classification}
\label{sub:classification}

To ensure that the temporal resolution of the light curves
is high enough to identify potential plateau features, we established a threshold for the signal-to-noise ratio of $\rm {S/N_{th}}=10$ and excluded all SGRBs with XRT data below this threshold. We find that 25 bursts out of 85 have a $\rm {S/N<S/N_{th}}$ and thus conclude that the available statistics for these GRBs were inadequate for a precise analysis of the light curve morphology. We refer to these GRBs as 'S/N-rejected'.

In one case, GRB 150101B\footnote{This burst was triggered with Fermi/Gamma-ray Burst Monitor
(Cummings et al., GCN 17267) and later on was found in the \textit{Swift}/BAT data during ground-based analysis (Cummings et al. GCN 17268).}, despite the $\rm {S/N>S/N_{th}}$,  the XRT follow-up started too late ($>1$day) to allow any identification of a plateau in its early afterglow.   
Consequently, we did not include it in the subsequent analysis.

\begin{figure}
                   \resizebox{\hsize}{!}{\includegraphics{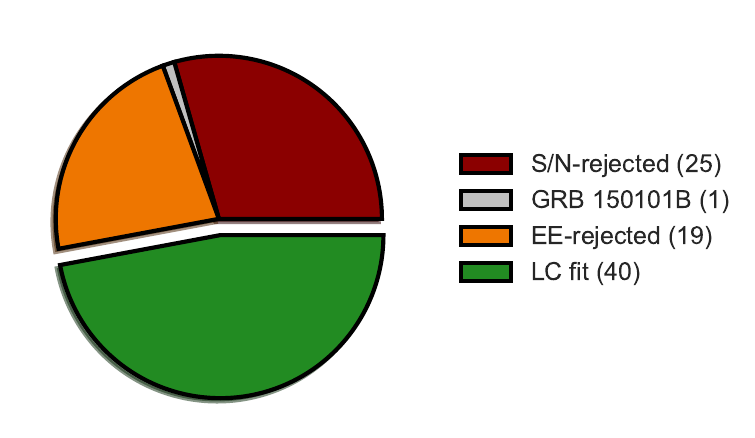}}
                        \caption{Pie chart showing the results of our preliminary classification of the 85 SGRBs included in the initial sample. Three subsamples were excluded as they were not suitable for the light curve analysis: S/N-rejected (25, in brown), EE-rejected (19, in orange), and GRB 150101B (in grey). The remaining 40 cases, defined as the LC fit sample (in green), are those deemed suitable for further analysis.}
                        \label{fig:rejected}
                \end{figure}

In 19 bursts with $\rm {S/N>S/N_{th}}$,
we find that the bulk emission is confined within the first few hundred seconds from the burst onset and its properties are better compatible with the so-called extended emission (EE) phase\footnote{This component can also be referred to as internal plateau.} rather than the afterglow. The EE is a prolonged prompt emission with softer properties with respect to the hard spike that characterises typical SGRBs, and it lasts a few hundred seconds \citep{norris2006, norris2010}. Only a fraction of SGRBs show an EE component, the origin of which is still debated and possibly connected with fallback accretion onto the central remnant \citep[e.g.][]{Rosswog2007,Kisaka_2015,Musolino2024ApJ...966L..31M}. To identify these 19 GRBs with EE features, we used the following general criteria:
            \begin{enumerate}
                \item The bulk of XRT data lie between $\sim$100 and $\sim$500 seconds from the trigger time.
                \item A pronounced temporal variability in the 
                photon index $\Gamma$  is present, typically showing a softening trend. 
                \item A rapid flux decay marks the EE phase end, with $F(t)\propto t^{\alpha}$ and typical $\alpha>2$ (sometimes $\gg2$)\footnote{In analogy with the steep decay phase observed in long GRBs  \citep[e.g.][]{Nousek2006ApJ...642..389N}, and interpreted as the tail of the prompt emission \citep{ZhangB_2006ApJ...642..354Z}.}.
            \end{enumerate}
In the 19 events identified with EE, the afterglow component is not detected or is too faint to allow any morphological study of the light curve. For these reasons, we also discarded this subsample from our analysis and labelled it `EE-rejected'.            
The remaining 40 SGRBs (47$\%$ of the initial sample) have enough statistics and sufficiently early temporal coverage in their X-ray afterglow light curves to allow us to identify plateaus if present. We label these events 
`LC fit' and we use this subsample to infer the fraction of SGRBs with evidence of a plateau in their afterglow light curve, as explained below.
The sample classification explained above is represented in a pie chart in Fig.~\ref{fig:rejected}.

\subsection{Afterglow light-curve morphology}
\label{sub:fit}
The analysis of the LC fit
sample of 40 SGRBs was performed 
by comparing the light curves with both a simple power-law and a broken power-law model. 

The simple power-law 
is written in the form:
\begin{equation}
F_{\mathrm{pl}}(t) = F_{\mathrm{pl,norm}}\left(\frac{t}{t_{\rm norm}}\right)^{-\alpha}
\label{eq:simplepl}
,\end{equation}
where $t_{\rm norm}$ is an arbitrary normalisation time, $F_{\mathrm{pl,norm}}$ is the normalisation constant (namely the value of the flux at $t_{\rm norm}$), and $\alpha$ is the power-law temporal decay index.
            
The smoothed broken power-law function (as presented in \citealt{Li_2012,Tang_2019}) is an 
empirical way to model a plateau phase,
and takes the form
\begin{equation}
F_{\mathrm{bpl}}(t) =  \frac{F_{\mathrm{break}}}{2^{-1/\omega}} \left[\left(\frac{t}{t_{\mathrm{break}}}\right)^{\alpha_1\omega} + \left(\frac{t}{t_{\mathrm{break}}}\right)^{\alpha_2\omega} \right]^{-1/\omega}\,,
\label{eq:bkpl}
\end{equation}
where $\alpha_1$ is the power-law index during the shallow phase (plateau),  $\alpha_2$ is the power-law index during the following decay phase, $t_{\mathrm{break}}$ is the 
power-law break time (observed end time of the plateau phase), $F_{\mathrm{break}}$ is the 
flux at $t_{\mathrm{break}}$,  
and $\omega$ is a smoothness parameter to control the sharpness of the transition between the plateau phase and the decay phase (high values correspond to  sharp breaks)
and is fixed to 
$\omega=3$ (e.g. \citealt{Li_2012, Yi_2016, Tang_2019}).

The reduced $\chi^2$ obtained from the fit performed with these two models were compared using an $F$-test to determine whether or not adding a temporal break to the simple power-law model leads to a statistically significant improvement. We used the 
threshold adopted in \citet{evans2009}, which means that we consider the cases that return an associated $p$-value of below the 4$\sigma$ threshold (corresponding to $6.2\times 10^{-5}$) as significant.

\begin{figure}
\centering
\begin{subfigure}{0.48\textwidth}
\centering
\resizebox{\hsize}{!}{\includegraphics{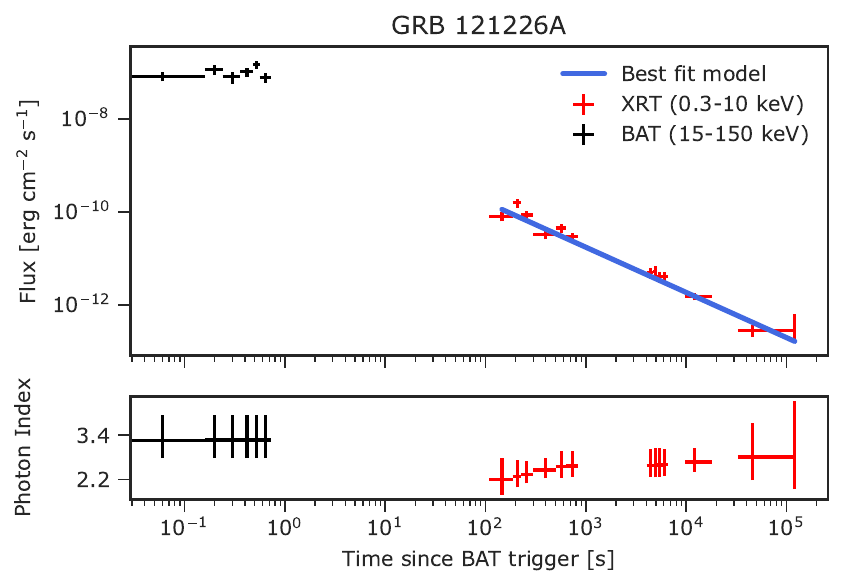}}
    \end{subfigure}
\begin{subfigure}{0.48\textwidth}
\centering
\resizebox{\hsize}{!}{\includegraphics{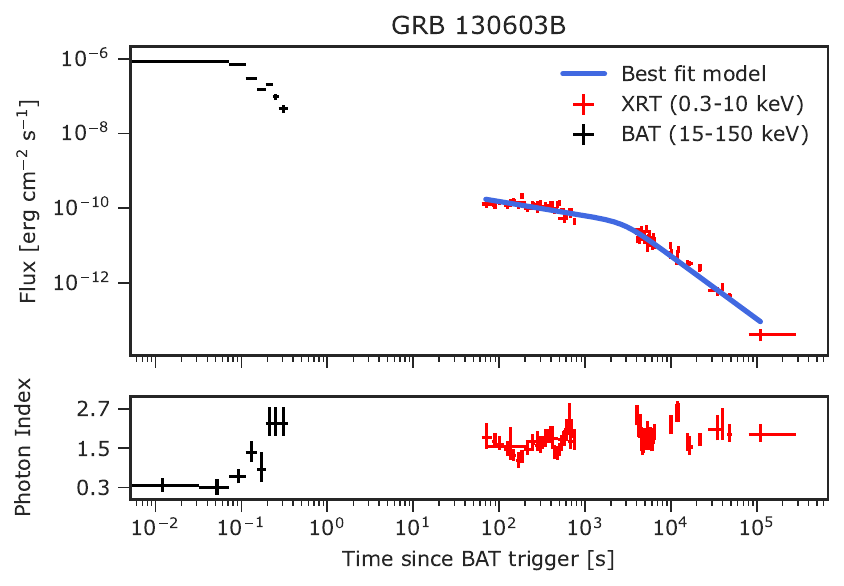}}
    \end{subfigure}
    \caption{
    {\it Swift} XRT (red) and BAT (black) light curves of two representative GRBs from the LC fit sample. 
    {\it Upper panel}:
    Example of a SGRB for which the fit with the simple power-law model (blue line) accurately reproduces the light curve behaviour.~{\it Lower panel}:~Example of a SGRB for which the broken power-law model (blue line) provides a more statistically significant fit.}
    \label{fig:lcs}
    \end{figure}

The results of the fit and of the $F$-test are reported in Table~\ref{tab:fit}. Two examples of a power-law and a broken power-law fit are presented in the upper and lower panels of Fig.~\ref{fig:lcs}, respectively. All the light curve fits are available on the GitHub repository dedicated to this work\footnote{\url{https://github.com/gugliluc/SGRB-thesis}}.
The analysis allowed us to identify 15 cases out of 40 (37.5$\%$) for which the break is found to be statistically significant: these define the `BPL' subsample. Conversely, for the other 25 cases (62.5$\%$), the power-law model is sufficient to describe the full flux time evolution: these bursts are thus identified as the `PL'
subsample.

Two of the cases included in the PL subsample, GRB 050724 and GRB 131004A, show strong flaring activity in the afterglow, the origin of which is still 
uncertain {and could be produced by either a black hole (BH) remnant
\citep{Perna2006,Proga2006,DallOsso2017} or a NS remnant \citep{Dai3,Dai2006}, or could be independent of the progenitor \citep{Giannios2006}}.
We discarded the flaring component 
in our analysis and prove that the behaviour of the remaining flux 
is well represented by a power-law.

\subsection{Incidence of plateaus in SGRBs}
\label{sec:incidence}
To identify the shallow phase of the 
BPL subsample with a plateau, we looked for the peculiar properties that characterise observed GRB plateaus, namely: (1) an anomalously shallow power-law decay index  $\alpha_1$  for the expected cooling regime of an afterglow at the observed epoch (typically $>0.5$ hours after the burst onset); and (2) no spectral variation around $t_{\mathrm{break}}$. 

Specifically, we checked if 
$\alpha_1$ is below the shallowest value predicted in a slow-cooling regime within the standard synchrotron scenario \citep[e.g.][]{sari1998}. By assuming a constant circumburst medium, $\alpha_1=3(1-p)/4\sim0.75$ for an electron energy power-law index $p\sim2$. 
This shallow decay is followed by a steepening accompanied by a softening of the spectrum. Indeed, in the slow-cooling regime, the shallowest light-curve power-law decay lasts until the electron cooling frequency $\nu_c$ drops below the observed band (0.3-10 keV in our case) and the synchrotron radiation spectral index increases from $(p-1)/2$ to $p/2$. Therefore, we also verified that no significant spectral softening is present  at the end of the plateau\footnote{In some specific conditions, for instance under the presence of a wind environment, values of $\alpha_1$ below 0.75 can be reached. However, the following steepening should be always accompanied by a softening behaviour.}.
    \begin{figure}
                \centering
                   \resizebox{\hsize}{!}{\includegraphics{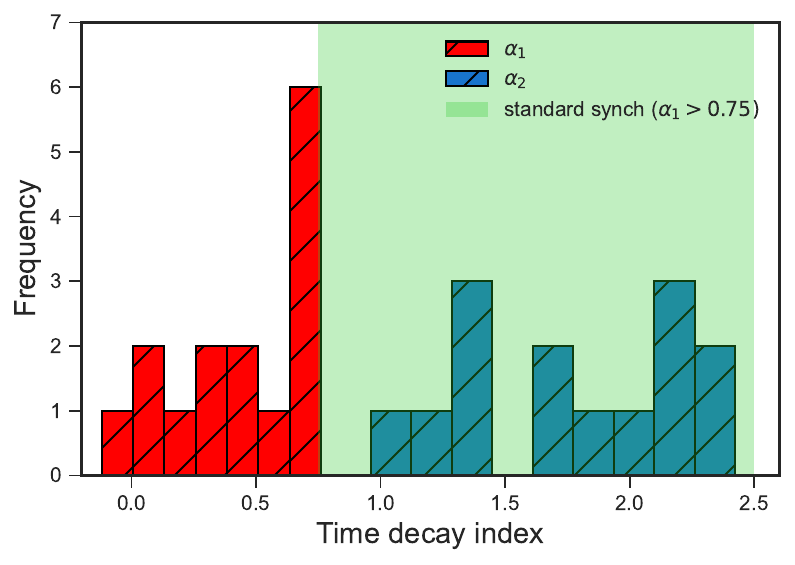}}
                        \caption{Power-law decay indices before ($\alpha_1$, in red) and after the break ($\alpha_2$, in blue) for the BPL subsample (see Sect.~\ref{sub:fit}). The green shaded area highlights the region that can be explained with the standard afterglow model (assuming the slow-cooling regime and a constant circumburst medium; e.g. \citealt{sari1998}), that is $0.75<\alpha_1<2.5$. 
          All the values we find for $\alpha_1$ cannot be explained with this model, suggesting the need for an alternative interpretation (see Sect.~\ref{sec:incidence} for further details).}
                        \label{fig:alpha1and2}
                \end{figure}
For all of the 15 SGRBs belonging to the BPL subsample, we find that the values of the decay index of the first power-law segment are $\alpha_1\le0.75$ (see Table~\ref{tab:fit}). This can be even more clearly seen in Fig.~\ref{fig:alpha1and2}, where the values of the indices before and after the break are represented in red and blue, respectively. The green shaded area indicates the values of $\alpha_1$ that can be explained within the standard synchrotron scenario. Additionally, we searched for spectral variability at the end of the plateau by extracting two spectra, one integrated during the plateau and another one during the post-plateau phase. By assuming a power-law spectral model, we find that the photon indexes are always compatible in the two phases.  
Despite the admittedly large uncertainties on the photon indexes, we conclude that there is no evidence of spectral variation for any of the events belonging to the BPL subsample.

These results imply that the standard synchrotron model fails to 
reproduce
the shallow phase and an alternative physical interpretation is required. The whole BPL subsample (15 events) could thus be classified as the `plateau' subsample, representing a fraction of SGRBs with evidence of an afterglow plateau of 37.5$\%$. 
If we assume that a plateau is absent not only in the 25 SGRBs forming the PL subsample but also in the 44 SGRBs excluded because of the low statistics in the afterglow component (25 S/N-rejected plus 19 EE-rejected), a lower limit to this fraction can then be obtained. If this is the case, 
the total sample increases from 40 to 
84, resulting in a plateau fraction of 15/84 (17.8$\%$). On the other hand, an upper limit can be given in the unlikely though not impossible scenario where a plateau is present in all of the 
44 faint SGRB afterglows. 
 In this case the plateau non-detection should be due to an intrinsic faintess and not to distance effect,
as the S/N-rejected subsample belongs to the most nearby SGRBs, with 23 of the 25 events below z=1, while 13 of the 19 EE-rejected events are found below z=1. In this scenario, the plateau fraction of SGRBs would account for the vast majority (70$\%$). 
However, an intrinsic faintness of the afterglow plateau is not only in contrast with the hypothesis whereby an energy injection mechanism is powering the early afterglow (independently of the origin of the energy injection) but is also disfavoured by past studies of the X-ray plateaus in long and SGRBs showing that the intrinsic plateau luminosity is well above the \textit{Swift}/XRT instrument detection limit \cite[see e.g.][Fig. 6]{Tang_2019}. We thus consider this scenario highly improbable, and the further considerations as to the energetics of GRBs with and without evidence of a plateau presented in the following section corroborate this conclusion.  

In conclusion, we estimate the incidence of SGRB plateaus ($f_{\rm pl}$ hereafter) to be within the range of 17.8\% to 37.5\%, and in all instances, it remains below the value observed for long GRBs  (>50\%).

\begin{figure}
\centering
\begin{subfigure}{0.48\textwidth}
\centering
\resizebox{\hsize}{!}{\includegraphics{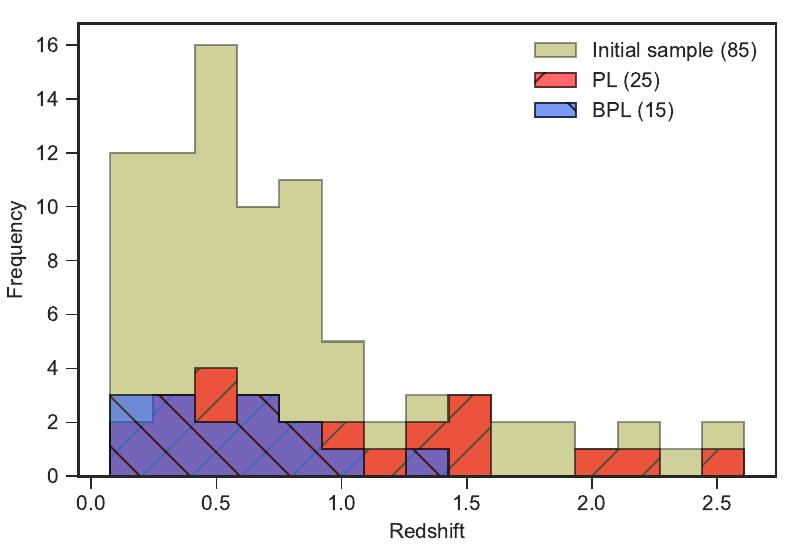}}
    \end{subfigure}
\begin{subfigure}{0.48\textwidth}
\centering
\resizebox{\hsize}{!}{\includegraphics{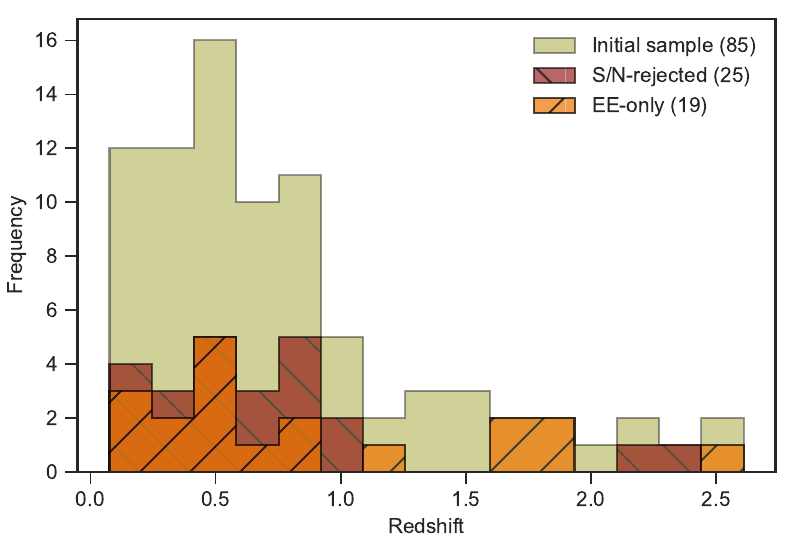}}
    \end{subfigure}
    \caption{
     Redshift distribution of the analyzed SGRBs (light green), compared to the distribution of  identified subsamples. 
    {\it Upper panel}:
    ~SGRBs with an X-ray afterglow light curve compatible with a broken power-law (blue) and a power-law (red) morphology.~{\it Lower panel}:~SGRBs not considered in our morphological analysis, 
    which are the S/N-rejected (in brown) and the EE-rejected (in orange) subsamples.}
    \label{fig:zdist}
    \end{figure}

\subsection{Testing for bias against distant SGRBs}
\label{sub:bias}

The upper panel of Fig. \ref{fig:zdist} shows the redshift distributions of the SGRBs with a plateau,  compared to SGRBs without a plateau and to all the events analyzed in this work.
SGRBs with a plateau are slightly skewed towards low redshifts, with no event with a plateau found above z$\sim$1.5, while several SGRBs with afterglow light curves compatible with simple power-law decay (i.e. no plateau) range up to z=2.5. 
To investigate any possible bias related to distance, 
we checked whether or not SGRBs with a plateau are also less energetic and therefore missed at high redshifts because they are too faint. To this aim, we collected the equivalent isotropic energy released during the burst, $E_{\rm iso}$, for each  SGRB of our sample for which this parameter was available in the literature (see Table~\ref{tab:eiso} for the complete list). When available, we took $E_{\rm iso}$ from the Konus-Wind (K-W) GRB catalogue (\citealt{tsvetkova_2017,tsvetkova_2021}), as the large spectral coverage of K-W allows us to reliably measure this parameter. We then compared the obtained $E_{\rm iso}$ distribution with that of the SGRBs belonging to the plateau subsample 
(Fig. \ref{fig:Eiso}). A visual inspection of the two distributions suggests that the SGRBs with plateaus have an $E_{\rm iso}$ distribution compatible with that of the total sample. 

At the same time, 
by considering the rejected subsample (i.e. S/N-rejected and EE-rejected subsamples), which are characterised by very poor statistics in the X-ray afterglow, we find no evidence for a distribution skewed towards high redshifts (lower panel of Fig. \ref{fig:zdist}), suggesting the presence of an intrinsically faint afterglow. As mentioned in the previous section, the presence of a faint plateau in these afterglows would be in contrast with the energy injection hypothesis ---regardless of the origin of the energy injection--- and with past studies showing plateau luminosity for long and SGRBs well above the \textit{Swift}/XRT detection threshold. The above considerations suggest that the results obtained in the previous sections are robust and strongly disfavour the presence of a plateau in the majority of SGRBs.

\section{Compatibility with the magnetar model}
\label{sec:magnetar}

The results presented in the previous section do not rely on any particular assumption regarding the origin of the plateau.~Here, we analyse those results in the context of the magnetar model ---that is assuming that the plateau is due to energy injection in the forward shock via magnetic dipole radiation from a millisecond-spinning magnetar formed in the BNS merger that produced the SGRB (see Sect.~\ref{sec:intro}).~We fit the plateau model presented in \citet{dallosso2011} and \citet{stratta2018} to afterglow light curves to estimate the spin period and dipole magnetic field of putative magnetars\footnote{\url{https://github.com/gistratta/magnetar}}.~Details of the model and of the fitting procedure are reported in Appendix \ref{app:magnetarmodel}.


\subsection{Magnetar model fitting}
\label{sub:magmod}
For the 15 SGRBs with evidence of an afterglow 
plateau, we calculated the 0.1-30 keV rest-frame emission as an approximation to the afterglow bolometric luminosity\footnote{This is strictly true when the photon index during the plateau and post-plateau is $< 2$, as verified in our sample, and provided that the cooling break is below a few times 30 keV.}, the determination of which would require more detailed knowledge of the intrinsic spectrum above 100 keV (rest-frame):
    \begin{equation}
       L(t) = 4\pi D_L^2(z)\times F_X(E_1, E_2, t)\times K_{\,[0.1-30 \,\mathrm{keV}]} \times  (1-\cos\theta_j),
    \label{eq:lumlc}
    \end{equation}
where $F_X$ is the observed X-ray flux in the $0.3-10$ keV band, $K$ is the cosmological correction,  
$f_b=(1-\cos\theta_j)$ is the jet beaming factor, and the observed  
time $t_{\mathrm{obs}}$ was converted to rest-frame time $t=t_{\mathrm{obs}}/(1+z)$. 

Measured jet half-opening angles were taken from the literature, where available (see Table~\ref{tab:angles}).~In cases with no  
 $\theta_j$ measurement, we adopted the values presented in \citet{zhu_2023}, based on the 
three-parameter correlation between $t_{\mathrm{jet},z}-E_{p,z}- E_{\mathrm{iso}}$, where $t_{\mathrm{jet},z}$ and $E_{p,z}$ are, respectively, the jet-break time and the peak energy in the source rest frame. In three cases, no estimate was provided even in \citet{zhu_2023}: for these cases, we assumed a fiducial value of $\theta_j=5,^\circ$ which is compatible with recent estimates for the majority of SGRBs \citep[e.g.][]{RoucoEscorial_2023ApJ...959...13R}.

\begin{table}
\centering  
\caption{Jet half-opening-angle values for the 15 SGRBs whose X-ray afterglow is compatible with the presence of a plateau (the plateau subsample).}
\begin{tabular}{lcccc}
\hline
\hline
GRB & $\theta_j$ & $\delta\theta_{j+}$ & $\delta\theta_{j-}$ & Reference\\
\hline
051221A  & 6       & 2.1   & 1.9   & 1 \\
060614         & 12.61   & 0.11  & 0.11  & 2 \\
061201         & 3.44    & 0.06  & 0.06  & 2 \\
070714B$^{p}$  & 8.59    & 0.92  & 0.69  & 2 \\
090510$^{p}$   & 2.29    & 0.11  & 0.11  & 2 \\
110402A        & 15.02   & 1.13  & 3.68  & 3 \\
130603B        & 6.3     & 1.7   & 5.1   & 1 \\
140903A        & 4       & 5     & 1.6   & 1\\
150424A        & 4.3     & 2.1   & 1.5   & 4  \\
151229A        & -       &   -   &   -   & - \\
161001A        & -       &   -   &   -   & - \\
170728B        & 3.5     & 1.1   & 0.8   & 4 \\
180618A        & -       &   -   &   -   & - \\
210323A$^{p}$  & 2.86    & 0.23  & 0.23  & 2 \\
211211A        & 6.86    & 0.12  & 0.12  & 2 \\
\hline
\label{tab:angles}
\end{tabular}
\tablebib{
   (1)\citet{aksulu_22}; (2) \citet{zhu_2023}; (3) \citet{Zhang_2015}; (4) \citet{RoucoEscorial_2023ApJ...959...13R}.
   }
\end{table}

\begin{figure}
    \centering
    \resizebox{\hsize}{!}{\includegraphics{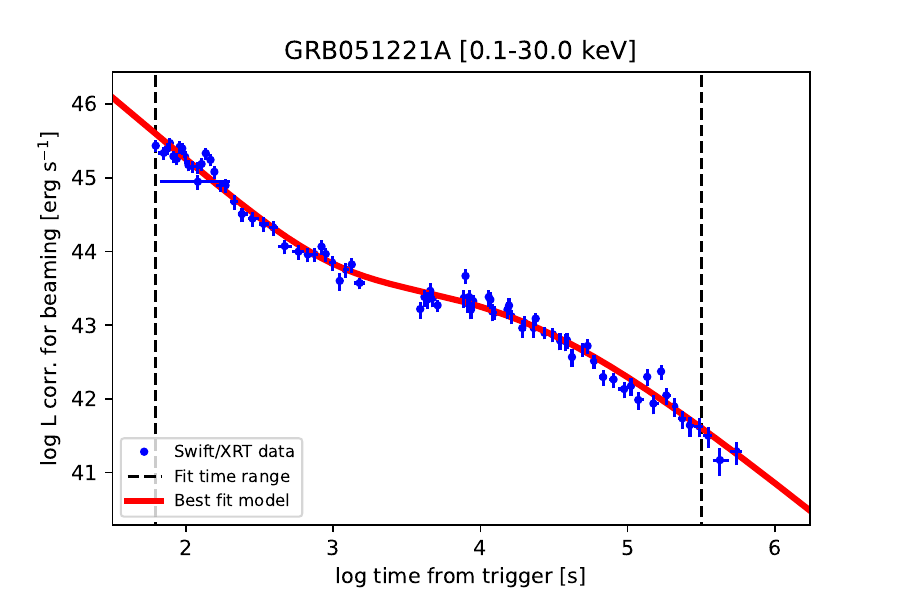}}
    \caption{Example of X-ray afterglow 0.1-30 keV rest-frame light curve (in blue) compared with the magnetar model (in red; see Sect.~\ref{sub:magmod}).}
    \label{fig:mag}
\end{figure}

The light curves of some bursts showed evidence of a steep decay phase before the plateau, as expected from the `canonical afterglow light curve' behaviour \citep{ZhangB_2006ApJ...642..354Z}.~To model this feature, we added an early power-law decay, with a free decay index, which allows us to better constrain the magnetar best-fit parameters from the plateau.~
In Fig.~\ref{fig:mag}, an example of a rest-frame light curve is presented, with the red line representing the magnetar model fit.
In Table~\ref{tab:magfit2} the best-fit values for the magnetar magnetic field, $B$, and spin period, $P$, are reported for the 13 bursts (out of 15) that gave good fits. All the magnetar model fits are available on the GitHub repository dedicated to this work\footnote{\url{https://github.com/gugliluc/SGRB-thesis}}.
For GRB 180618A, it was not possible to constrain the values of~$B$ and $P$ due to the very short duration of its plateau, which is significantly shorter than any of the others  (Table~\ref{tab:fit}).~This event was not classified as an EE because its post-plateau decay index is consistent with being $\lesssim 2$, and therefore does not fulfill our third criterion for EEs. However, along with its proximity to the prompt phase, it also shows evidence for a spectral softening ---followed by a moderate hardening--- between 300 and 600 s, thus satisfying the other two criteria for selecting EEs.~We propose that this short plateau may be best understood as an $\sim 10^2$ s EE, which was followed at $t > 10^3$ s by a standard afterglow with no plateau, thus explaining the failure of the magnetar model fit. 


In GRB 061201, on the other hand, the 
extremely large magnetic field $B>5.6\times 10^{16}$~G is close to the theoretical maximum, that is, the virial limit (e.g. \citealt{Reis09, Akg13}), and is therefore physically implausible. Moreover,
the very long spin period of $P >38$ ms also implies an 
exceedingly slow rotation for a NS formed in a BNS merger.~Also in this case, the magnetar model does not appear to provide a viable explanation, despite the formally acceptable fit.  
In conclusion, both GRB 180618A and GRB 061201 were rejected and included in the `failed magnetar' subsample.


\subsection{Considerations related to the `rejected' subsamples}
\label{subsec:criteria}

To further extend our analysis, we reconsidered the EE-rejected subsample in light of the magnetar model.~In these GRBs, which were rejected despite the high S/N of the XRT data, the spectral and temporal properties of the  light curve are consistent with EE only, with no afterglow.~We thus aimed to determine an upper limit to the 
luminosity of 
a possibly 
undetected afterglow plateau. By comparing with expectations of the magnetar model, we were able to rule out the presence of a magnetar central engine in a number of these GRBs.  

Our analysis relied on the following two general assumptions, which allow us to set a conservative minimum luminosity for a magnetar plateau, $L_{p, {\rm min}}$:
    \begin{itemize}
        \item 
        Driven by observations of plateaus in short and long GRBs, we took $\sim 10^5$~s as an estimate of the maximum plateau duration.
        \item 
        We adopted a very conservative upper limit of 30 ms for the spin period of a stable NS formed in a BNS merger, which corresponds to a rotational energy of $\sim 3 \times 10^{49}$ erg.
    \end{itemize}     
Based on these assumptions, the minimum luminosity of a magnetar plateau is
\begin{equation}
        L_p (t) \gtrsim L_{p,\mathrm{th}}\sim 3 \times 10^{44} \,\mathrm{erg}\, \mathrm{s}^{-1}\, ,
        \label{eq:lmin}
\end{equation}
which is simply based on the magnetic dipole formula, relating the plateau luminosity and duration to the initial spin energy of the NS (e.g. \citealt{dallossoEstella2022} and references therein).~By computing the luminosity corresponding to the last detection at the end of the EE for each burst as $L_{p,\mathrm{min}}$, all the EE-rejected GRBs
with $L_{p, {\rm min}} < L_{p, {\rm th}}$ were classified as inconsistent with the presence of a magnetar plateau\footnote{Since we assume an isotropically emitting magnetar, we calculated the isotropic-equivalent luminosities  from the data to compare with $L_{p,\mathrm{th}}$. }.~We call this the $L_{p, {\rm min}}$-criterion.
We note that a similar minimum luminosity can also be derived in an alternative scenario where plateaus are the results of an off-axis viewing angle towards a structured jet (\citealt{Beniamini_2020}; Fig.~2). 

The second criterion is 
based on the scenario in which a newly born magnetar generates the GRB prompt emission during an early accretion phase (see \citealt{Dall_Osso_2023}). A minimum accretion luminosity is achieved during the prompt phase, at which the magnetar enters the propeller regime and the luminosity starts a steady and steep decline.~In this framework, the ratio $\kappa$ is defined 
between the intrinsic (i.e. corrected for the jet beaming factor) minimum luminosity in the prompt and the initial spin-down luminosity of the magnetar, that is the isotropic-equivalent luminosity of the plateau.
Interestingly, $\kappa$ is set primarily by the magnetar spin period and the NS mass and radius (see eq.~5 in \citealt{Dall_Osso_2023}); therefore, it cannot be arbitrarily large. 
~In particular, for an assumed maximum spin period of 30~ms, one obtains $\kappa \lesssim 30$. 
To estimate the beamed-corrected minimum prompt luminosity for the 19 GRBs in the 
EE-rejected 
subsample, 
we adopted a common value 
$f_b \sim 0.01$ of the beaming factor (see Appendix \ref{app:Lmin}).~All cases in which $\kappa >30$ are thus inconsistent with the presence of a magnetar plateau.~We label this as the $\kappa$-criterion. 

Combining these two criteria, we were able to rule out the presence of a magnetar plateau in 9 out of 19 bursts.  
The remaining 10 were renamed `EE-rejected (inconclusive)'.
\subsection{Magnetar fraction}
\label{sub:magnetarfraction}
By including the 13 good 'magnetar'
candidates (see Sect. \ref{sub:magmod}) with respect to the initial sample of 85 bursts, the associated magnetar fraction is $f_{\mathrm{mag}}^{\rm min}=0.152$. In contrast, barring all cases for which no conclusion could be drawn on the presence of a plateau (i.e. the 25 S/N-rejected subsample, GRB 150101B, and the 10 EE-rejected (inconclusive) GRBs), the magnetar fraction is computed over a total of $49$ bursts,  
leading to $f_{\mathrm{mag}}^{\rm max}=0.265$.~We conclude that the fraction of SGRBs that are fully consistent with a magnetar central engine is in the range of $15.2\%$ to $26.5\%$.

\section{Astrophysical implications}
\label{sec:maxmasstop}

\begin{table*}
\centering
\caption{Best-fit values of the magnetar magnetic field strength $B$ and initial spin period $P$ for the 13 out of 15 cases for which the magnetar model fit of the luminosity light curve is successful.
}

\begin{tabular}{lccccccc}
\hline
\hline
\multicolumn{1}{l}{GRB name} &\multicolumn{2}{c}{Input}& \multicolumn{4}{c}{Output}\\
\cmidrule(lr){2-3}\cmidrule(lr){4-8}
\multicolumn{1}{l}{}& $z$    & $\theta_j$ & $B$ & $P$  & $\tau_{\rm sd} \approx 680~\left(\displaystyle\frac{P_{\rm ms}}{B_{15}}\right)^2$  &$\chi^2$ & $\nu$ \\
\multicolumn{1}{l}{}&     & (deg) &  ($10^{14}$ G) & (ms)  & (ks) & &  \\
\hline
051221A & 0.5464 &  6.0    &   $(29\pm2)$          &$(12.8\pm0.3)$  & 13.3 $\pm$ 2.1 & 125   & 77  \\
060614  & 0.125  &  12.6&  $(37\pm3)$          &$(24\pm1)$ & $34\pm9$& 1749  & 465 \\
070714B & 0.923  &  8.6 &   $(132\pm40)$        &$(11\pm2)$ & 0.5 $\pm0.3$& 284   & 79  \\
090510   & 0.903  &  2.3 &   $(82\pm7)$         &$(4.5\pm0.2)$ & $0.20 \pm 0.04$ & 80   & 63  \\
110402A& 0.854  &  15.0&   $(96\pm37)$        &$(14\pm1)$ & 1.4 $\pm1.1$ & 42.0   & 16  \\
130603B  & 0.3568 &  6.3  &   $(110\pm2)$        &$(13.2\pm0.2)$  & 0.98 $\pm0.04$ & 137   & 70  \\
140903A & 0.3529 &  4.0    &   $(32\pm4)$          &$(8\pm0.3)$ & 4.5 $\pm1.2$ & 56   & 36  \\
150424A& 0.3 & 4.3 &  $(36\pm4)$          &$(16\pm1)$ & 13 $\pm3$ & 243    & 115 \\
151229A  & 0.63    &  5.0  & (67$\pm$9) & (4.3$\pm0.9$) & 0.3 $\pm0.1$ & 142    & 56  \\ 
161001A & 0.67&  5.0 &  $(47\pm6)$          &$(4.2\pm0.2)$ & $0.5\pm0.1$ & 88    & 54  \\
170728B  & 1.272  &  3.5    & $(20\pm1)$      &$(1.50\pm0.03)$ & 0.39 $\pm0.04$ & 232   & 193 \\
210323A & 0.733  &  2.9 & $(51\pm11)$          &$(8.7\pm0.6)$ & 2.0 $\pm0.9$ & 62    & 18  \\
211211A & 0.0763 & 6.9   & $(286\pm26)$        &$(27\pm1)$ & 0.6 $\pm0.1$ & 779  & 265  \\
\hline
\label{tab:magfit2}
\end{tabular}
\tablefoot{Besides the redshift $z$ and the jet half-opening angle $\theta_j$ of each SGRB, the table also shows the spin-down timescale $\tau_{sd}$ implied by $B$ and $P$ (see e.g. \citealt{dallossoEstella2022}), as well as the fit $\chi^2$ and degrees of freedom $\nu$.}
\end{table*}

\subsection{Outcomes of BNS mergers}
\label{sec:maxmass}
Three main outcomes are expected for BNS merger 
remnants (e.g. \citealt{Faber_2012, Baio17, Piro_2017, Margalit_2017, Rezzolla_2018, Bern20}).~If 
the masses of the two NSs are low enough, a stable NS remnant can form with a mass 
lower than the Tolman-Oppenheimer-Volkoff mass ($M_{\rm TOV}$), the maximum NS stable mass of non-rotating configurations \citep{Oppenheimer_PhysRev.55.374}.~A~fast rotation of the remnant, which may be inherited from the angular momentum of the binary, 
can delay the NS collapse into a BH
for an extended time\footnote{Typically $\sim$~tens-to-hundreds of seconds (e.g. \citealt{lasky14, dall15}), until spindown losses, such as GW emission or magnetic dipole radiation, significantly reduce the~rotation.} (supra-massive NS), even if its mass is larger than $M_{\rm TOV}$ yet below the maximum $M_{\rm max}\approx 1.2 M_{\rm TOV}$ (e.g. \citealt{Baum00, Brez16, Marg22,Musolino2024ApJ...962...61M}).~Above $M_{\rm max}$, prompt collapse to a BH is generally expected\footnote{Hyper-massive NSs may form, which are briefly ($\lesssim$ hundreds of milliseconds) sustained by short-lived differential rotation.}.~Thus, the fate of the remnant is strongly dependent on its mass and on $M_{\rm TOV}$; 
 this latter in turn depends on the NS equation of state (EoS). 

Independently of the NS EoS, a key assumption for the considerations that follow is that, if a stable (or supra-massive) NS is formed, it should have a 
very strong magnetic field. Indeed, for a 
wide range of 
EoS and initial magnetic fields of the merging NSs, general relativistic magnetohydrodynamic (MHD) simulations by a variety of groups have robustly demonstrated that, due to a combination of magnetic winding from differential rotation, and hydrodynamic and MHD instabilities, the magnetic field is amplified to magnetar levels \citep{Giaco11, Giacomazzo2013,Giacomazzo2015,Kiuchi2015,Ciolfi2017,Kiuchi2018,Ciolfi2019,Palenzuela2022,Aguilera2023,Kiuchi2024}.
Therefore, we can confidently assume in our analysis that, for those NS-NS mergers that leave behind a stable or long-lived NS, this remnant NS will have the magnetic field typical of a magnetar.

Another important assumption that is key to the considerations that follow is the ability of a NS-NS merger to lead to an ultrarelativistic jet (as observed for GRB170817; e.g. \citealt{Mooley_2018,Lazzati2018,Ioka2018,Alexander2018}), independently of the type of merger remnant. Recent MHD simulations by \citet{Bamber2024} with low-mass NSs have indeed shown that jet-like structures, satisfying their criteria for an incipient jet, were observed also for the supra-massive NS remnants. However, the question of whether a relativistic jet could be launched and break out from the ejecta could not be fully answered and needs to wait for MHD simulations to extend to much larger scales. With this caveat in mind, the following results are built on the assumption that a SGRB can be produced also by a stable or supra-massive, highly magnetised NS remnant.
  \begin{figure*}
\centering
\begin{subfigure}{0.48\textwidth}
\centering
\resizebox{\hsize}{!}{\includegraphics{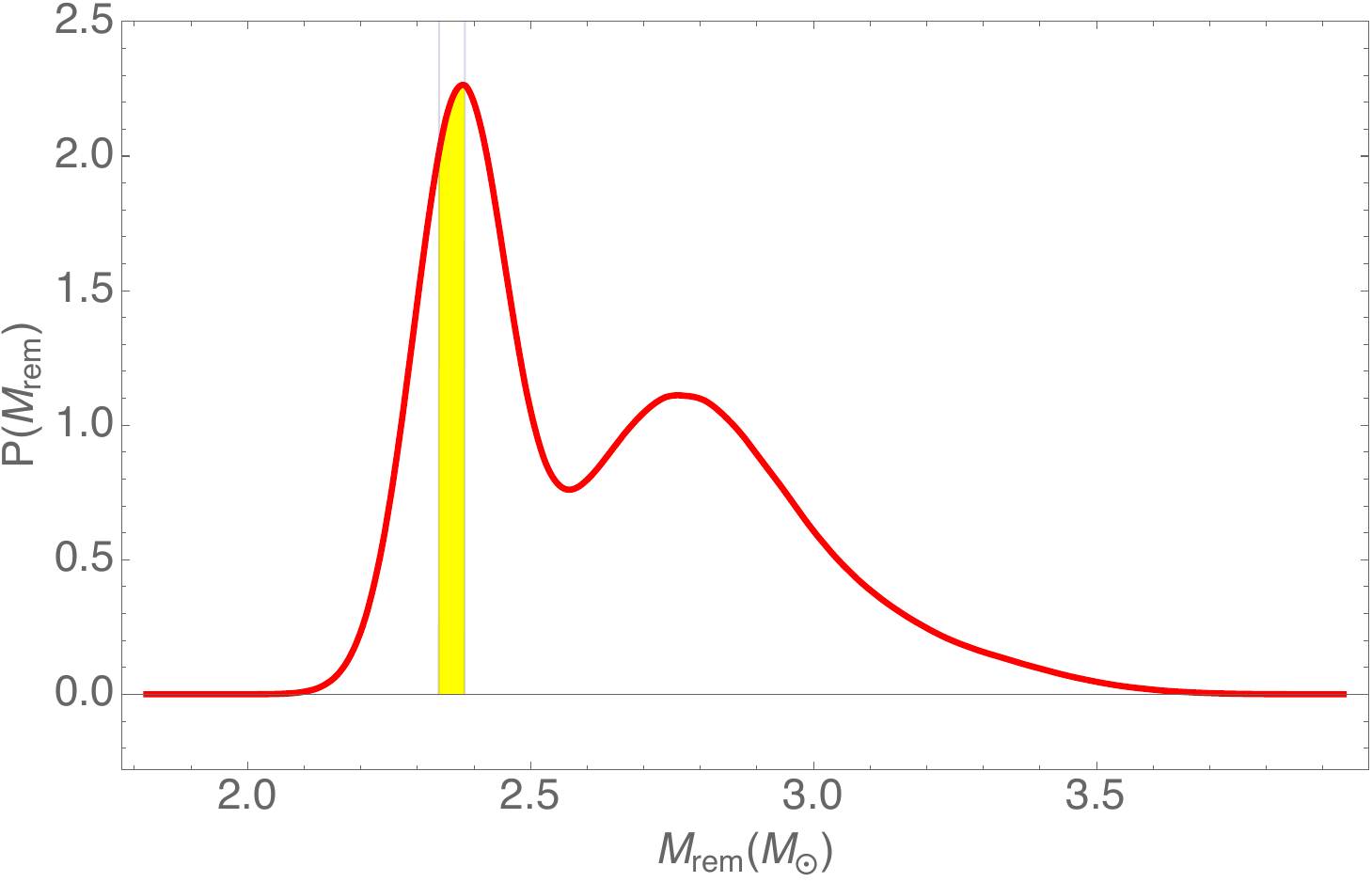}}
    \end{subfigure}
\begin{subfigure}{0.48\textwidth}
\centering
\resizebox{\hsize}{!}{\includegraphics{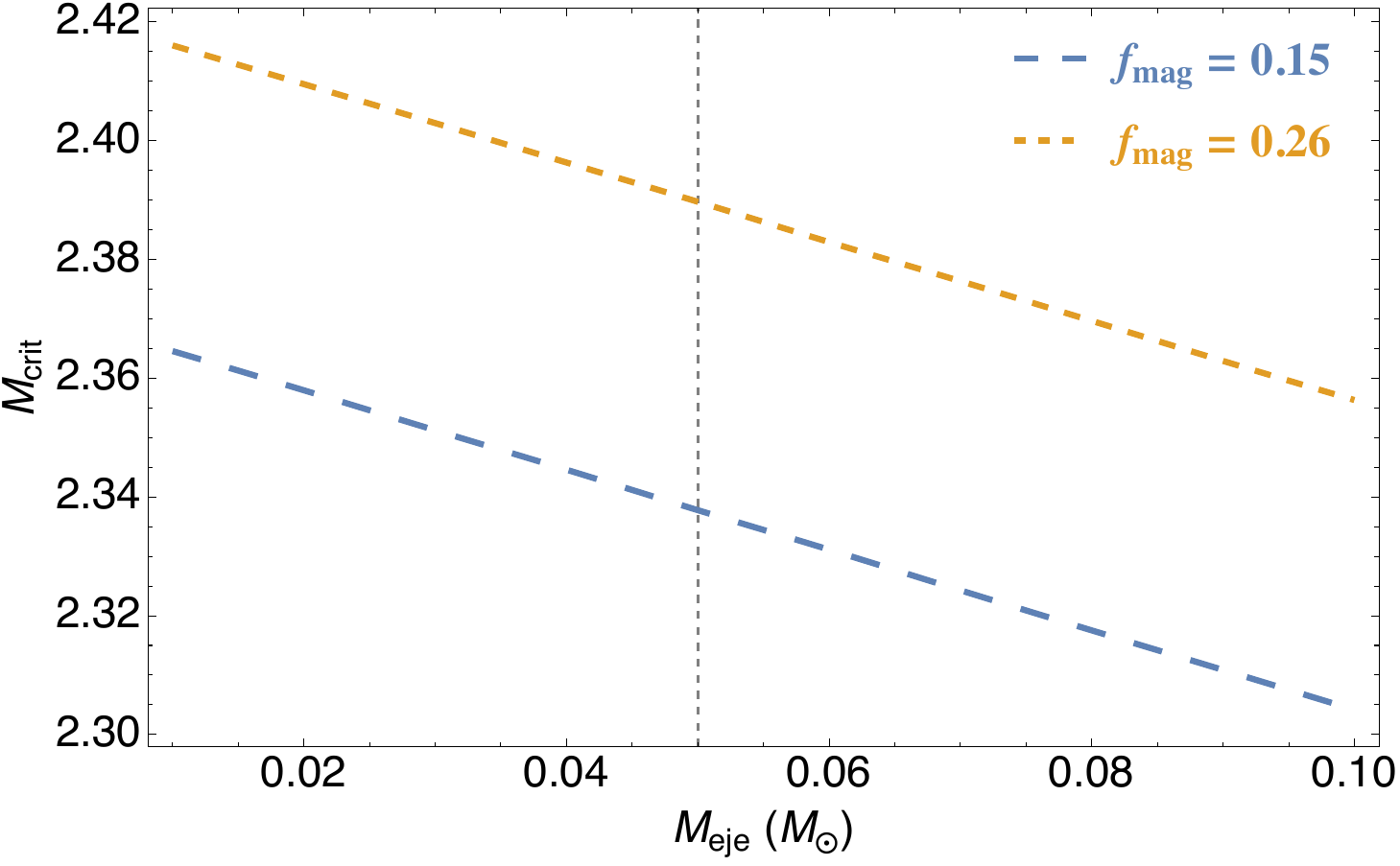}}
    \end{subfigure}
    \caption{
     The mass probability distribution of merger remnants ({\it left panel}) and the range of $M_{\rm crit}$ as a function of the adopted value for the ejecta mass, $M_{\rm ej}$ ({\it right panel}).
    The mass distribution of merger remnants is obtained by assuming (i) a double-peaked Gaussian mass distribution for binary NS components, with one peak at $\mu_1 = 1.34$ M$_\odot$ ($\sigma_1$ = 0.07 M$_\odot$) and one at $\mu_2=1.8$ M$_\odot$ ($\sigma_2$ = 0.21 M$_\odot$), (ii) a quasi-EOS-independent relation between gravitational and baryonic mass (Eq.~\ref{eq:lattimer}), and (iii) a fiducial ejecta mass of $\sim$ 0.05 M$_\odot$.~The yellow-shaded area indicates the $M_{\rm rem}$ range for which the cumulative distribution function yields a probability of $\sim $0.15-0.26 (i.e. our $f_{\rm mag}$ range).~It defines our estimated range for $M_{\rm crit}$, that is the maximum mass of NS remnants that remain stable or do not collapse into a BH for at least a time long enough to power plateaus lasting $>$ 200 s, such as those found in our sample.}
    \label{fig:mtov}
    \end{figure*}

In this scenario, the fraction of SGRBs with evidence of plateaus is directly linked to the fraction of BNSs that form a stable NS remnant or a supra-massive NS that does not collapse to a BH at least for the duration of the observed plateau. 
Within this framework, 
we can thus set constraints 
on the remnant mass as the critical value $M_{\rm crit}$ at which its cumulative distribution returns the same plateau fraction value, as detailed in the following section.

\subsection{Estimating $M_{\rm crit}$} 
\label{sub:mcrit}
We first computed the probability distribution of the merger remnant mass starting from the distribution of the  mass of the two components. 
Simulations of massive star 
 explosions indicate~a bimodal NS mass distribution \citep{ZhangW2008ApJ...679..639Z}.~In the present~work,
 the gravitational masses of the two binary components, $M_{g,1}$ and $M_{g,2}$, were drawn from the preferred double-peaked Gaussian distribution for NSs found in binary systems in our Galaxy, with means $\mu_1 = 1.34\,$M$_{\odot}$ and $\mu_2 = 1.8\,$M$_{\odot}$, and standard deviations $\sigma_1 = 0.07\,$M$_{\odot}$ and $\sigma_2 = 0.21\,$M$_{\odot}$ (\citealt{Alsing18,Farr20, Rocha23}). 


~The 
baryonic mass of the remnants was calculated using the quasi-EOS-independent relation between the gravitational ($M_g$) and baryonic ($M_b$) mass of NSs (\citealt{lattimer_2021}): 
 \begin{equation}
 \label{eq:lattimer}
M_b = M_g + \left(0.062\pm 0.0016\right) M^2_g +\left(0.018\pm 0.0065\right) M^3_g \, .
 \end{equation}

 ~In addition to the masses of the two components, we further accounted for the mass ejected during the merger ($M_{\rm ej}$): given the large uncertainties on this quantity as derived by different numerical simulations, we assumed   
a `fiducial' value of $M_{\rm ej} = 0.05$ M$_\odot$ in all mergers, motivated by the ejecta mass in GW 170817 (e.g. 
\citealt{Coughlin2018}).


The baryonic mass distribution of the remnants is given by
  \begin{equation}
            M_{b,\mathrm{rem}} = M_{b,1} + M_{b,2} - M_{\mathrm{ej}},
          \label{eq:mb_remn}
 \end{equation}
 which is 
 then converted into gravitational mass $M_g$ according to Eq.~\ref{eq:lattimer}.
 We finally compute the 
 critical mass ($M_{\rm crit}$) at which the remnant mass cumulative distribution function returns a value equal to the magnetar fraction, $f_{\mathrm{mag}}$.~The left panel of Fig.~\ref{fig:mtov} 
 presents our results when adopting a 'fiducial' value of $M_{\mathrm{ej}}=0.05$~M$_\odot$, and for $f_{\mathrm{mag}} =0.15-0.26$, which define the yellow-shaded confidence region in the plot.~From the latter, we obtain $M_{\rm crit} \sim (2.34-2.39)$~M$_\odot$.~The right panel of Fig.~\ref{fig:mtov} illustrates how this range depends on $M_{\rm ej}$:~we conclude that, for any plausible value of the ejecta mass, we have $ 2.31 < M_{\rm crit}/{\rm M}_{\odot}< 2.41$.~Finally, we also tried a different distribution for NS masses in the form of a single Gaussian with mean $\mu =1.34 {\rm M}_\odot$ and $\sigma=0.06 {\rm M}_\odot$ (\citealt{ozel_2016}):~in this case, the allowed interval for $M_{\rm crit}$ is reduced by 0.1 M$_\odot$ (i.e. $2.2 \lesssim M_{\rm crit}/{\rm M}_\odot \lesssim 2.3$) for the same $M_{\rm ej}$ values.

If all the plateaus in our sample were produced by stable NSs, then
$M_{\rm crit}$ 
would be equal to  
$M_{\rm TOV}$,  
while the former would be larger than the latter if some plateaus were associated with supra-massive NSs that later collapsed to BHs.~Thus, the condition 
$M_{\rm TOV}\leq M_{\rm crit}< M_{\rm max} \approx 1.2M_{\rm TOV}$ 
holds in general, as for masses above 
$M_{\rm max}$
a merger remnant promptly collapses
to a BH.

\section{Discussion}

\label{sec:discussion}

\subsection{Origin of the plateau}
In the above analysis, 
we adopt an empirical approach~to quantify 
the incidence of observed plateaus ($f_{\rm pl}$) in SGRBs 
with no assumptions as to their origin, obtaining  $0.18<~f_{\rm pl} <~0.37$.~An 
important implication of this result is that~any theoretical interpretation of afterglow plateaus should 
not~only account for this observed $f_{\rm pl}$ in SGRBs, but should also explain 
why it appears to be inconsistent with (significantly lower than) the  $f_{\rm pl}$  of long GRBs \citep[e.g.][]{Tang_2019}. 

In the `structured jet model' \cite[e.g.][]{Beniamini_2020}, the different values of $f_{\rm pl}$ in short and long GRBs impose the existence~of two~significantly different jet structures, with highly constrained parameters within each class (see also \citealt{Ocon24}).
 While possible, this may also be in tension with the results of numerical simulations, which show significant variations in jet structure even within a single GRB class (e.g. \citealt{LazPEr19, aksulu_22, Urru21, Garcia24}). 
 On the other hand, 
the `magnetar model' provides a more straightforward interpretation of our results. The magnetar population associated with core-collapse supernovae, which represents $\sim$ 10\% of the NS population (\citealt{gaensler2005}) or possibly even more (\citealt{Beniamini_2019}), can in principle account for a sizeable fraction of plateaus observed in long GRBs.~ Conversely, as noted above
only a small fraction of BNS mergers will produce a magnetar remnant, and therefore a plateau.

We note that, in principle, a fraction $f'$ of SGRBs may originate from NS-BH mergers, from which only a BH remnant could form. In this case, the fraction of SGRBs that is  associated with NS remnants decreases by a factor $(1-f')$, and the true $f_{\rm pl}$, as well as 
$M_{\rm crit}$, should be increased by $1/(1-f')$.~However, since the fraction of SGRBs from NS-BH mergers is still uncertain, and is thought to be much smaller than that from BNSs 
\citep[e.g.][]{Foucart2012PhRvD..86l4007F},  we expect our estimate of 
the critical mass to remain largely unaffected.

\subsection{Constraining $M_{\rm TOV}$}
\label{sub:mtovcon}

Our main conclusion is that 13 
SGRBs 
in our sample show evidence of a plateau that can be well modelled~as due to energy injection from a magnetar central engine.~This~corresponds to a magnetar fraction of 
$\sim$ 15-26\%, depending on the cut on the total sample, as discussed in Sects.~\ref{sec:incidence} and \ref{subsec:criteria}.~The latter provides a realistic estimate of~$f_{\rm mag}$, as we have ruled out selection biases that~might~cause plateaus~to~be over-represented in the 46 discarded afterglows. Adopting a fiducial value of $M_{\rm ej} = 0.05$ M$_\odot$ for the ejecta mass, this range of $f_{\rm mag}$ constrains the critical mass, $M_{\rm crit} \approx (2.34-2.39)$~M$_\odot$, up~to~which NS remnants can survive long enough to power the observed plateaus.~Moreover, to check the dependence of our result on $M_{\rm ej}$, we allowed the latter to vary in the range (0-0.1)~M$_\odot$, determining a corresponding range of $2.31<M_{\mathrm{crit}}/{\rm M}_{\odot}<2.41$, as depicted in the right panel of Fig.~\ref{fig:mtov}.

\begin{figure}
    \centering
    \resizebox{\hsize}{!}{\includegraphics{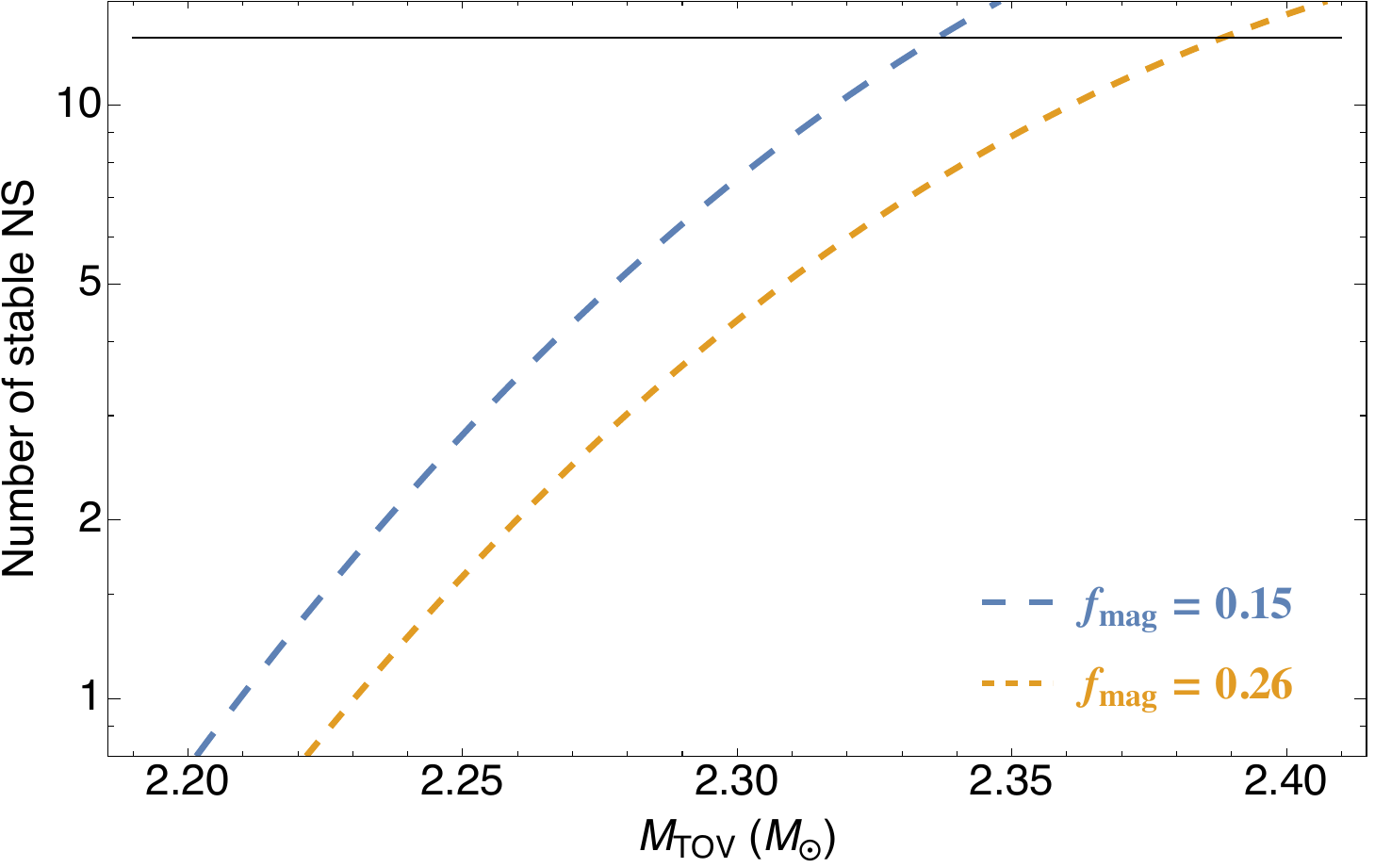}}
    \caption{Number of stable or long-lived NSs in our sample as a function of $M_{\rm TOV}$.~The two dashed curves correspond to the minimum and maximum fractions found in this work.~The upper, horizontal black curve indicates the total of 13 NSs.}
    \label{fig:mtovlim}
\end{figure}

The number of indefinitely stable NSs in our plateau sample depends on $M_{\rm TOV}$; the lower this latter value is, the larger the fraction of supra-massive NSs, and the faster their collapse due to the increasing importance of rotational support 
in holding these NSs from collapsing into BHs.
~As~a consequence, the relatively long duration ($> $ a few ks) of most plateaus in our sample and the relatively slow NS spins (see Table~\ref{tab:magfit2}) point to a significant fraction of NSs being indefinitely stable\footnote{Or so marginally above $M_{\rm TOV}$ that they effectively track its value to a good accuracy (e.g. \citealt{dall15} and references therein).}.~Using the probability distribution of remnant masses (left panel of Fig.~\ref{fig:mtov}), we calculated the number of stable or long-lived (supra-massive) NSs expected in our sample as a function of $M_{\rm TOV}$, and report our results in Fig.~\ref{fig:mtovlim}.

Additionally, a lower limit to $M_{\rm TOV}$ can be placed by assuming that the fastest-rotating magnetar in our sample (GRB 170728B) was also the 
one with the closest $M$  to $M_{\rm crit}$.~With~a nominal initial spin period of $\sim 1.5$~ms (Table~\ref{tab:magfit2}), this NS may have collapsed at a spin period of $P_{\rm coll} \gtrsim 2.1$~ms (beyond the end of the plateau):~adopting this spin at collapse, along with  
numerical approximations for the critical mass as a function of angular momentum, and for other relevant quantities of relativistic rotating NSs (\citealt{Brez16, Musolino2024ApJ...962...61M}), we estimate a lower limit to $M_{\rm TOV}$ of  $\gtrsim 2.35(2.3) {\rm M}_\odot$ for $f_{\rm mag} = 0.26(0.15)$, and a spin parameter of $a_{\rm crit} =J/M_{\rm crit}^2 \approx 0.25$, roughly 35\% of the Keplerian limit.~It is interesting to note that, for this $M_{\rm TOV}$, Fig.~\ref{fig:mtovlim} indicates that 8-9 stable NSs should be expected, out of the 13 in our sample.~Indeed, at least 8 GRBs with relatively slowly rotating magnetars (spin period $> 5$ ms) and long plateaus ($>$ a few ks) can be identified in Table~\ref{tab:magfit2}, likely indicating stable NSs with negligible centrifugal support.

\subsection{Comparing with previous results}
\label{sub:comparison}
Several authors have exploited the rich observations of GW~170817 and GRB 170817A to place significant constraints on $M_{\rm TOV}$ based on the lack of signatures of the presence of a long-lived remnant in that merger (e.g. \citealt{Margalit_2017, Bauswein_2017, Rezzolla_2018}).~Broadband data are indeed consistent with, and indicative of, the formation of a hypermassive NS, which collapsed to a BH within less than a second after the merger (e.g. \citealt{shibata2019}). 

Under the assumption that the remnant was formed with nearly maximal rotation, inherited from the orbital angular momentum of the progenitor binary, the inevitable implication of its early collapse is that its mass 
was very~close~to $M_{\rm max}\approx~1.2 M_{\rm TOV}$.~Given the accurate mass measurements for both the binary components and the ejecta mass, values of $M_{\rm TOV} < 2.2 {\rm M}_\odot$ were thus derived.

Our result, on the other hand, points to a different scenario, in which $M_{\rm TOV} \gtrsim 2.3$M$_\odot$ and merger remnants can form with significantly sub-Keplerian rotation, as indicated by the inferred spin periods of our fits.~Thanks to the large $M_{\rm TOV}$, remnants with masses of $\lesssim 2.4$ M$_\odot$ can be sustained against collapse even with a modest centrifugal support
for a time long enough to power plateaus that last $>$ ks.~This is 
consistent with the  conclusions of \cite{Marg22} that merger remnants may form with slow rotation, and that most of the angular momentum is deposited in the surrounding torus, from which it is progressively accreted.~If the initial mass of the slowly rotating remnant is large enough, a small amount of accreted mass may lead to collapse before achieving a significant spin-up.~Within this framework, \cite{Marg22} concluded that the broadband data from GRB 170817A are consistent with $M_{\rm TOV} \sim (2.3-2.5)$M$_\odot$ if, at collapse, the merger remnant was rotating at 0-0.5 times the Keplerian rate.

\section{Summary and conclusions}
\label{sec:conclusion}
In this study, we analysed the most comprehensive sample~of SGRBs with known redshifts observed by the \textit{Swift} X-ray telescope  to date, intending to establish the fraction of events displaying a plateau phase in their X-ray afterglows.~Among the 85 events in our sample, only 40 allowed for morphological analysis of the light curve and for a spectral evolution study (Sect.~\ref{sub:classification}).~
We find 15 afterglows that show evidence of a plateau (Sect.~\ref{sec:incidence}); this is roughly one-third of the total ($f_{\rm pl}\sim 0.18-0.37$) and significantly less than the fraction observed in long GRBs (> 50\%; e.g. \citealt{Tang_2019}).~This difference represents a first important constraint for plateau models.

By fitting the 15 plateau light curves with a model of energy injection from a spinning down magnetar, we find very good fits in 13 cases;~further analysis of the two outliers suggests the misclassification of an extended prompt phase (GRB 180618A), or a different origin for the plateau feature (GRB 061201).~The 13 acceptable fits define our `true' magnetar sample.~Moreover, for 9 additional GRBs showing an extended emission (EE) in their prompt phase with faint or no afterglow, we rule out the presence of a plateau. Indeed, for these 9 events, we note that the
large luminosity drops at the end of the EE (Sect.~\ref{subsec:criteria}) 
is in general inconsistent with known correlations between prompt and plateau luminosities 
(e.g. \citealt{Dainotti2016ApJ...825L..20D}, \citealt{Beniamini_2020}, 
\citealt{Dall_Osso_2023}).~As a consequence, the number of SGRBs for which we are able to infer the presence or absence of a plateau rises to 49, 
implying a magnetar fraction of $f_{\rm mag} \sim 0.15-0.26$ in our sample (either referring to the total of 85 events or to the subsample of 49 valid GRBs; Sect.~\ref{sub:magnetarfraction}).
 
Under the assumption that all (or at least the vast majority of) SGRBs originate from BNS mergers, we used $f_{\rm mag}$ to infer the maximum mass ($M_{\rm crit}$) of magnetars powering the observed plateaus.~Adopting a double-peaked Gaussian for the mass distribution of NSs in binaries, we derive $M_{\rm crit}~\approx (2.31-2.41)$M$_\odot$ (varying the ejecta mass from 0.1 M$_\odot$ to 0, see Fig.~\ref{fig:mtov}), where in general  $M_{\rm TOV} \le M_{\rm crit} < M_{\rm max} \approx 1.2 M_{\rm TOV}$ (Sect.~\ref{sub:mcrit}).~If all magnetars born~in~BNS mergers were long-lived, then $M_{\rm crit} \approx~M_{\rm TOV}$.~If, on the other hand, all such magnetars were supra-massive objects sustained against collapse by fast rotation then  $M_{\rm crit} \lesssim ~M_{\rm max}$, and hence $M_{\rm TOV} \gtrsim (1.93-2) {\rm M}_\odot$. In Fig.~\ref{fig:mtovlim} we illustrate the relation between the number of stable or long-lived magnetars and the implied $M_{\rm TOV}$.

Our fits suggest that most plateau-powering magnetars are born with relatively slow rotation ($P > 5$ ms) and can survive for relatively long times ($>$ ks);~thus, we conclude that, in most cases, a long-lived or indefinitely stable NS was formed, and
 the computed value of $M_{\rm crit}$ is thus expected to be close to, yet slightly above, $M_{\rm TOV}$.~Taking a further step, we placed a lower limit on $M_{\rm TOV} \gtrsim 2.35(2.3) {\rm M}_\odot$ for $f_{\rm mag} =0.26(0.15)$ by making the reasonable assumption that the fastest-rotating magnetar in our sample is also the one closest to $M_{\rm crit}$, and that it collapsed to a BH 
within a few hundred seconds of its birth (Sect.~\ref{sub:mtovcon}).~These 
conclusions align with the latest findings by \cite{Fan3} and with the scenario recently proposed by \cite{Marg22} regarding the formation and early evolution of BNS merger remnants.~They are, at the same time, in tension with previous estimates of $M_{\rm TOV}$ derived from observations of GW 170817/GRB 170817A, and based on the assumption that the BNS remnant was formed with near-maximal rotation (Sect.~\ref{sub:comparison}).

Looking ahead, future observations of SGRBs, such as those expected with the ESA space mission project THESEUS (e.g. \citealt{Amati_2018, Stratta_2018}), in synergy with the third-generation gravitational wave interferometers, such as the Einstein Telescope (e.g. \citealt{Punturo_2010}), hold promise for providing direct evidence of the magnetar hypothesis for afterglow plateaus via detection of the faint, transient modulated gravitational wave signal expected shortly after the formation of a NS remnant 
(e.g. \citealt{dallossoEstella2022} and references therein).

\begin{acknowledgements} 
We thank Prof. Luciano Rezzolla for his insightful comments and careful reading of the manuscript.~G.S. acknowledges the support by the State of Hesse within the Research Cluster ELEMENTS (Project ID 500/10.006). 
R.P. gratefully acknowledges support by NSF award AST-2006839.~S.D. acknowledges funding by the European Union’s~Horizon2020 research and innovation programme under the Marie Skłodowska-Curie (grant agreement No.754496), and the support from the Giersch Science Center at the Goethe University of Frankf\"{u}rt, where part this work was carried out.~This work made use of data supplied by the UK \textit{Swift} Science Data Centre at the University of Leicester. 
\end{acknowledgements}

\bibliographystyle{aa} 
\bibliography{main} 

\begin{thebibliography}{104}
\expandafter\ifx\csname natexlab\endcsname\relax\def\natexlab#1{#1}\fi

\bibitem[{{Aguilera-Miret} {et~al.}(2023){Aguilera-Miret}, {Palenzuela}, {Carrasco}, \& {Vigan{\`o}}}]{Aguilera2023}
{Aguilera-Miret}, R., {Palenzuela}, C., {Carrasco}, F., \& {Vigan{\`o}}, D. 2023, \prd, 108, 103001

\bibitem[{{Akg{\"u}n} {et~al.}(2013){Akg{\"u}n}, {Reisenegger}, {Mastrano}, \& {Marchant}}]{Akg13}
{Akg{\"u}n}, T., {Reisenegger}, A., {Mastrano}, A., \& {Marchant}, P. 2013, \mnras, 433, 2445

\bibitem[{Aksulu {et~al.}(2022)Aksulu, Wijers, vanEerten, \& van~der Horst}]{aksulu_22}
Aksulu, M.~D., Wijers, R. A. M.~J., vanEerten, H.~J., \& van~der Horst, A.~J. 2022, \mnras, 511, 2848

\bibitem[{{Alexander} {et~al.}(2018){Alexander}, {Margutti}, {Blanchard}, {Fong}, {Berger}, {Hajela}, {Eftekhari}, {Chornock}, {Cowperthwaite}, {Giannios}, {Guidorzi}, {Kathirgamaraju}, {MacFadyen}, {Metzger}, {Nicholl}, {Sironi}, {Villar}, {Williams}, {Xie}, \& {Zrake}}]{Alexander2018}
{Alexander}, K.~D., {Margutti}, R., {Blanchard}, P.~K., {et~al.} 2018, \apjl, 863, L18

\bibitem[{{Alsing} {et~al.}(2018){Alsing}, {Silva}, \& {Berti}}]{Alsing18}
{Alsing}, J., {Silva}, H.~O., \& {Berti}, E. 2018, \mnras, 478, 1377

\bibitem[{Amati {et~al.}(2018)Amati, O’Brien, Götz, Bozzo, Tenzer, Frontera, Ghirlanda, Labanti, Osborne, Stratta, Tanvir, Willingale, Attina, Campana, Castro-Tirado, Contini, Fuschino, Gomboc, Hudec, Orleanski, Renotte, Rodic, Bagoly, Blain, Callanan, Covino, Ferrara, Le~Floch, Marisaldi, Mereghetti, Rosati, Vacchi, D’Avanzo, Giommi, Piranomonte, Piro, Reglero, Rossi, Santangelo, Salvaterra, Tagliaferri, Vergani, Vinciguerra, Briggs, Campolongo, Ciolfi, Connaughton, Cordier, Morelli, Orlandini, Adami, Argan, Atteia, Auricchio, Balazs, Baldazzi, Basa, Basak, Bellutti, Bernardini, Bertuccio, Braga, Branchesi, Brandt, Brocato, Budtz-Jorgensen, Bulgarelli, Burderi, Camp, Capozziello, Caruana, Casella, Cenko, Chardonnet, Ciardi, Colafrancesco, Dainotti, D’Elia, De~Martino, De~Pasquale, Del~Monte, Della~Valle, Drago, Evangelista, Feroci, Finelli, Fiorini, Fynbo, Gal-Yam, Gendre, Ghisellini, Grado, Guidorzi, Hafizi, Hanlon, Hjorth, Izzo, Kiss, Kumar, Kuvvetli, Lavagna, Li, Longo, Lyutikov, Maio, Maiorano,
  Malcovati, Malesani, Margutti, Martin-Carrillo, Masetti, McBreen, Mignani, Morgante, Mundell, Nargaard-Nielsen, Nicastro, Palazzi, Paltani, Panessa, Pareschi, Pe’er, Penacchioni, Pian, Piedipalumbo, Piran, Rauw, Razzano, Read, Rezzolla, Romano, Ruffini, Savaglio, Sguera, Schady, Skidmore, Song, Stanway, Starling, Topinka, Troja, van Putten, Vanzella, Vercellone, Wilson-Hodge, Yonetoku, Zampa, Zampa, Zhang, Zhang, Zhang, Zhang, Antonelli, Bianco, Boci, Boer, Botticella, Boulade, Butler, Campana, Capitanio, Celotti, Chen, Colpi, Comastri, Cuby, Dadina, De~Luca, Dong, Ettori, Gandhi, Geza, Greiner, Guiriec, Harms, Hernanz, Hornstrup, Hutchinson, Israel, Jonker, Kaneko, Kawai, Wiersema, Korpela, Lebrun, Lu, MacFadyen, Malaguti, Maraschi, Melandri, Modjaz, Morris, Omodei, Paizis, Páta, Petrosian, Rachevski, Rhoads, Ryde, Sabau-Graziati, Shigehiro, Sims, Soomin, Szécsi, Urata, Uslenghi, Valenziano, Vianello, Vojtech, Watson, \& Zicha}]{Amati_2018}
Amati, L., O’Brien, P., Götz, D., {et~al.} 2018, Advances in Space Research, 62, 191–244

\bibitem[{{Baiotti} \& {Rezzolla}(2017)}]{Baio17}
{Baiotti}, L. \& {Rezzolla}, L. 2017, Reports on Progress in Physics, 80, 096901

\bibitem[{Bamber {et~al.}(2024)Bamber, Tsokaros, Ruiz, \& Shapiro}]{Bamber2024}
Bamber, J., Tsokaros, A., Ruiz, M., \& Shapiro, S.~L. 2024, \prd, 110

\bibitem[{{Baumgarte} {et~al.}(2000){Baumgarte}, {Shapiro}, \& {Shibata}}]{Baum00}
{Baumgarte}, T.~W., {Shapiro}, S.~L., \& {Shibata}, M. 2000, \apjl, 528, L29

\bibitem[{Bauswein {et~al.}(2017)Bauswein, Just, Janka, \& Stergioulas}]{Bauswein_2017}
Bauswein, A., Just, O., Janka, H.-T., \& Stergioulas, N. 2017, \apjl, 850, L34

\bibitem[{Beniamini {et~al.}(2022)Beniamini, Gill, \& Granot}]{Beniamini_2022}
Beniamini, P., Gill, R., \& Granot, J. 2022, \mnras, 515, 555

\bibitem[{Beniamini {et~al.}(2020)Beniamini, Granot, \& Gill}]{Beniamini_2020}
Beniamini, P., Granot, J., \& Gill, R. 2020, \mnras, 493, 3521–3534

\bibitem[{Beniamini {et~al.}(2019)Beniamini, Hotokezaka, van~der Horst, \& Kouveliotou}]{Beniamini_2019}
Beniamini, P., Hotokezaka, K., van~der Horst, A., \& Kouveliotou, C. 2019, \mnras, 487, 1426

\bibitem[{{Bernuzzi}(2020)}]{Bern20}
{Bernuzzi}, S. 2020, General Relativity and Gravitation, 52, 108

\bibitem[{{Breu} \& {Rezzolla}(2016)}]{Brez16}
{Breu}, C. \& {Rezzolla}, L. 2016, \mnras, 459, 646

\bibitem[{{Ciolfi} {et~al.}(2017){Ciolfi}, {Kastaun}, {Giacomazzo}, {Endrizzi}, {Siegel}, \& {Perna}}]{Ciolfi2017}
{Ciolfi}, R., {Kastaun}, W., {Giacomazzo}, B., {et~al.} 2017, \prd, 95, 063016

\bibitem[{{Ciolfi} {et~al.}(2019){Ciolfi}, {Kastaun}, {Kalinani}, \& {Giacomazzo}}]{Ciolfi2019}
{Ciolfi}, R., {Kastaun}, W., {Kalinani}, J.~V., \& {Giacomazzo}, B. 2019, \prd, 100, 023005

\bibitem[{{Costa} {et~al.}(1997){Costa}, {Frontera}, {Heise}, {Feroci}, {in't Zand}, {Fiore}, {Cinti}, {Dal Fiume}, {Nicastro}, {Orlandini}, {Palazzi}, {Rapisarda\#}, {Zavattini}, {Jager}, {Parmar}, {Owens}, {Molendi}, {Cusumano}, {Maccarone}, {Giarrusso}, {Coletta}, {Antonelli}, {Giommi}, {Muller}, {Piro}, \& {Butler}}]{costa1997}
{Costa}, E., {Frontera}, F., {Heise}, J., {et~al.} 1997, \nat, 387, 783

\bibitem[{{Coughlin} {et~al.}(2018){Coughlin}, {Dietrich}, {Doctor}, {Kasen}, {Coughlin}, {Jerkstrand}, {Leloudas}, {McBrien}, {Metzger}, {O'Shaughnessy}, \& {Smartt}}]{Coughlin2018}
{Coughlin}, M.~W., {Dietrich}, T., {Doctor}, Z., {et~al.} 2018, \mnras, 480, 3871

\bibitem[{{Dai}(2004)}]{Dai3}
{Dai}, Z.~G. 2004, \apj, 606, 1000

\bibitem[{{Dai} \& {Lu}(1998{\natexlab{a}})}]{Dai2}
{Dai}, Z.~G. \& {Lu}, T. 1998{\natexlab{a}}, \prl, 81, 4301

\bibitem[{{Dai} \& {Lu}(1998{\natexlab{b}})}]{Dai1}
{Dai}, Z.~G. \& {Lu}, T. 1998{\natexlab{b}}, \aap, 333, L87

\bibitem[{{Dai} {et~al.}(2006){Dai}, {Wang}, {Wu}, \& {Zhang}}]{Dai2006}
{Dai}, Z.~G., {Wang}, X.~Y., {Wu}, X.~F., \& {Zhang}, B. 2006, Science, 311, 1127

\bibitem[{{Dainotti} {et~al.}(2016){Dainotti}, {Postnikov}, {Hernandez}, \& {Ostrowski}}]{Dainotti2016ApJ...825L..20D}
{Dainotti}, M.~G., {Postnikov}, S., {Hernandez}, X., \& {Ostrowski}, M. 2016, \apjl, 825, L20

\bibitem[{{Dall'Osso} {et~al.}(2015){Dall'Osso}, {Giacomazzo}, {Perna}, \& {Stella}}]{dall15}
{Dall'Osso}, S., {Giacomazzo}, B., {Perna}, R., \& {Stella}, L. 2015, \apj, 798, 25

\bibitem[{{Dall'Osso} {et~al.}(2017){Dall'Osso}, {Perna}, {Tanaka}, \& {Margutti}}]{DallOsso2017}
{Dall'Osso}, S., {Perna}, R., {Tanaka}, T.~L., \& {Margutti}, R. 2017, \mnras, 464, 4399

\bibitem[{{Dall'Osso} \& {Stella}(2022)}]{dallossoEstella2022}
{Dall'Osso}, S. \& {Stella}, L. 2022, in \apss, Vol. 465, \apss, ed. S.~{Bhattacharyya}, A.~{Papitto}, \& D.~{Bhattacharya}, 245--280

\bibitem[{Dall'Osso {et~al.}(2023)Dall'Osso, Stratta, Perna, Cesare, \& Stella}]{Dall_Osso_2023}
Dall'Osso, S., Stratta, G., Perna, R., Cesare, G.~D., \& Stella, L. 2023, \apjl, 949, L32

\bibitem[{Dall’Osso {et~al.}(2011)Dall’Osso, Stratta, Guetta, Covino, De~Cesare, \& Stella}]{dallosso2011}
Dall’Osso, S., Stratta, G., Guetta, D., {et~al.} 2011, \aap, 526, A121

\bibitem[{{Dereli-B{\'e}gu{\'e}} {et~al.}(2022){Dereli-B{\'e}gu{\'e}}, {Pe'er}, {Ryde}, {Oates}, {Zhang}, \& {Dainotti}}]{Dereli2022NatCo..13.5611D}
{Dereli-B{\'e}gu{\'e}}, H., {Pe'er}, A., {Ryde}, F., {et~al.} 2022, Nature Communications, 13, 5611

\bibitem[{{Evans} {et~al.}(2009){Evans}, {Beardmore}, {Page}, {Osborne}, {O'Brien}, {Willingale}, {Starling}, {Burrows}, {Godet}, {Vetere}, {Racusin}, {Goad}, {Wiersema}, {Angelini}, {Capalbi}, {Chincarini}, {Gehrels}, {Kennea}, {Margutti}, {Morris}, {Mountford}, {Pagani}, {Perri}, {Romano}, \& {Tanvir}}]{evans2009}
{Evans}, P.~A., {Beardmore}, A.~P., {Page}, K.~L., {et~al.} 2009, \mnras, 397, 1177

\bibitem[{{Evans} {et~al.}(2007){Evans}, {Beardmore}, {Page}, {Tyler}, {Osborne}, {Goad}, {O'Brien}, {Vetere}, {Racusin}, {Morris}, {Burrows}, {Capalbi}, {Perri}, {Gehrels}, \& {Romano}}]{evans2007}
{Evans}, P.~A., {Beardmore}, A.~P., {Page}, K.~L., {et~al.} 2007, \aap, 469, 379

\bibitem[{Faber \& Rasio(2012)}]{Faber_2012}
Faber, J.~A. \& Rasio, F.~A. 2012, Living Reviews in Relativity, 15

\bibitem[{{Fan} {et~al.}(2024){Fan}, {Han}, {Jiang}, {Shao}, \& {Tang}}]{Fan3}
{Fan}, Y.-Z., {Han}, M.-Z., {Jiang}, J.-L., {Shao}, D.-S., \& {Tang}, S.-P. 2024, \prd, 109, 043052

\bibitem[{{Farr} \& {Chatziioannou}(2020)}]{Farr20}
{Farr}, W.~M. \& {Chatziioannou}, K. 2020, Research Notes of the American Astronomical Society, 4, 65

\bibitem[{{Fong} {et~al.}(2015){Fong}, {Berger}, {Margutti}, \& {Zauderer}}]{fong_2015}
{Fong}, W., {Berger}, E., {Margutti}, R., \& {Zauderer}, B.~A. 2015, \apj, 815, 102

\bibitem[{{Fong} {et~al.}(2022){Fong}, {Nugent}, {Dong}, {Berger}, {Paterson}, {Chornock}, {Levan}, {Blanchard}, {Alexander}, {Andrews}, {Cobb}, {Cucchiara}, {Fox}, {Fryer}, {Gordon}, {Kilpatrick}, {Lunnan}, {Margutti}, {Miller}, {Milne}, {Nicholl}, {Perley}, {Rastinejad}, {Escorial}, {Schroeder}, {Smith}, {Tanvir}, \& {Terreran}}]{fong2022}
{Fong}, W.-f., {Nugent}, A.~E., {Dong}, Y., {et~al.} 2022, \apj, 940, 56

\bibitem[{{Foucart}(2012)}]{Foucart2012PhRvD..86l4007F}
{Foucart}, F. 2012, \prd, 86, 124007

\bibitem[{{Gaensler} {et~al.}(2005){Gaensler}, {McClure-Griffiths}, {Oey}, {Haverkorn}, {Dickey}, \& {Green}}]{gaensler2005}
{Gaensler}, B.~M., {McClure-Griffiths}, N.~M., {Oey}, M.~S., {et~al.} 2005, \apjl, 620, L95

\bibitem[{{Gao} \& {Fan}(2006)}]{Fan1}
{Gao}, W.-H. \& {Fan}, Y.-Z. 2006, \cjaa, 6, 513

\bibitem[{{Garc{\'\i}a-Garc{\'\i}a} {et~al.}(2024){Garc{\'\i}a-Garc{\'\i}a}, {L{\'o}pez-C{\'a}mara}, \& {Lazzati}}]{Garcia24}
{Garc{\'\i}a-Garc{\'\i}a}, L., {L{\'o}pez-C{\'a}mara}, D., \& {Lazzati}, D. 2024, \mnras, 531, 2903

\bibitem[{{Gehrels} {et~al.}(2004){Gehrels}, {Chincarini}, {Giommi}, {Mason}, {Nousek}, {Wells}, {White}, {Barthelmy}, {Burrows}, {Cominsky}, {Hurley}, {Marshall}, {M{\'e}sz{\'a}ros}, {Roming}, {Angelini}, {Barbier}, {Belloni}, {Campana}, {Caraveo}, {Chester}, {Citterio}, {Cline}, {Cropper}, {Cummings}, {Dean}, {Feigelson}, {Fenimore}, {Frail}, {Fruchter}, {Garmire}, {Gendreau}, {Ghisellini}, {Greiner}, {Hill}, {Hunsberger}, {Krimm}, {Kulkarni}, {Kumar}, {Lebrun}, {Lloyd-Ronning}, {Markwardt}, {Mattson}, {Mushotzky}, {Norris}, {Osborne}, {Paczynski}, {Palmer}, {Park}, {Parsons}, {Paul}, {Rees}, {Reynolds}, {Rhoads}, {Sasseen}, {Schaefer}, {Short}, {Smale}, {Smith}, {Stella}, {Tagliaferri}, {Takahashi}, {Tashiro}, {Townsley}, {Tueller}, {Turner}, {Vietri}, {Voges}, {Ward}, {Willingale}, {Zerbi}, \& {Zhang}}]{gehrelsetal2004}
{Gehrels}, N., {Chincarini}, G., {Giommi}, P., {et~al.} 2004, \apj, 611, 1005

\bibitem[{{Giacomazzo} \& {Perna}(2013)}]{Giacomazzo2013}
{Giacomazzo}, B. \& {Perna}, R. 2013, \apjl, 771, L26

\bibitem[{{Giacomazzo} {et~al.}(2011){Giacomazzo}, {Rezzolla}, \& {Baiotti}}]{Giaco11}
{Giacomazzo}, B., {Rezzolla}, L., \& {Baiotti}, L. 2011, \prd, 83, 044014

\bibitem[{{Giacomazzo} {et~al.}(2015){Giacomazzo}, {Zrake}, {Duffell}, {MacFadyen}, \& {Perna}}]{Giacomazzo2015}
{Giacomazzo}, B., {Zrake}, J., {Duffell}, P.~C., {MacFadyen}, A.~I., \& {Perna}, R. 2015, \apj, 809, 39

\bibitem[{{Giannios}(2006)}]{Giannios2006}
{Giannios}, D. 2006, \aap, 455, L5

\bibitem[{{Ioka} \& {Nakamura}(2018)}]{Ioka2018}
{Ioka}, K. \& {Nakamura}, T. 2018, Progress of Theoretical and Experimental Physics, 2018, 043E02

\bibitem[{Kisaka \& Ioka(2015)}]{Kisaka_2015}
Kisaka, S. \& Ioka, K. 2015, \apjl, 804, L16

\bibitem[{{Kiuchi} {et~al.}(2015){Kiuchi}, {Cerd{\'a}-Dur{\'a}n}, {Kyutoku}, {Sekiguchi}, \& {Shibata}}]{Kiuchi2015}
{Kiuchi}, K., {Cerd{\'a}-Dur{\'a}n}, P., {Kyutoku}, K., {Sekiguchi}, Y., \& {Shibata}, M. 2015, \prd, 92, 124034

\bibitem[{{Kiuchi} {et~al.}(2018){Kiuchi}, {Kyutoku}, {Sekiguchi}, \& {Shibata}}]{Kiuchi2018}
{Kiuchi}, K., {Kyutoku}, K., {Sekiguchi}, Y., \& {Shibata}, M. 2018, \prd, 97, 124039

\bibitem[{{Kiuchi} {et~al.}(2024){Kiuchi}, {Reboul-Salze}, {Shibata}, \& {Sekiguchi}}]{Kiuchi2024}
{Kiuchi}, K., {Reboul-Salze}, A., {Shibata}, M., \& {Sekiguchi}, Y. 2024, Nature Astronomy, 8, 298

\bibitem[{{Lasky} {et~al.}(2014){Lasky}, {Haskell}, {Ravi}, {Howell}, \& {Coward}}]{lasky14}
{Lasky}, P.~D., {Haskell}, B., {Ravi}, V., {Howell}, E.~J., \& {Coward}, D.~M. 2014, \prd, 89, 047302

\bibitem[{Lattimer(2021)}]{lattimer_2021}
Lattimer, J. 2021, Annual Review of Nuclear and Particle Science, 71, 433

\bibitem[{Lattimer \& Prakash(2007)}]{LATTIMER_2007}
Lattimer, J. \& Prakash, M. 2007, \physrep, 442, 109

\bibitem[{{Lazzati} \& {Perna}(2019)}]{LazPEr19}
{Lazzati}, D. \& {Perna}, R. 2019, \apj, 881, 89

\bibitem[{Lazzati {et~al.}(2023)Lazzati, Perna, Gompertz, \& Levan}]{Lazzati_2023}
Lazzati, D., Perna, R., Gompertz, B.~P., \& Levan, A.~J. 2023, \apjl, 950, L20

\bibitem[{{Lazzati} {et~al.}(2018){Lazzati}, {Perna}, {Morsony}, {Lopez-Camara}, {Cantiello}, {Ciolfi}, {Giacomazzo}, \& {Workman}}]{Lazzati2018}
{Lazzati}, D., {Perna}, R., {Morsony}, B.~J., {et~al.} 2018, \prl, 120, 241103

\bibitem[{Li {et~al.}(2012)Li, Liang, Tang, Chen, Xi, Lü, Gao, Zhang, Zhang, Yi, Lu, Lü, \& Wei}]{Li_2012}
Li, L., Liang, E.-W., Tang, Q.-W., {et~al.} 2012, \apj, 758, 27

\bibitem[{{Margalit} {et~al.}(2022){Margalit}, {Jermyn}, {Metzger}, {Roberts}, \& {Quataert}}]{Marg22}
{Margalit}, B., {Jermyn}, A.~S., {Metzger}, B.~D., {Roberts}, L.~F., \& {Quataert}, E. 2022, \apj, 939, 51

\bibitem[{Margalit \& Metzger(2017)}]{Margalit_2017}
Margalit, B. \& Metzger, B.~D. 2017, \apjl, 850, L19

\bibitem[{Metzger {et~al.}(2011)Metzger, Giannios, Thompson, Bucciantini, \& Quataert}]{Metzger_2007}
Metzger, B.~D., Giannios, D., Thompson, T.~A., Bucciantini, N., \& Quataert, E. 2011, \mnras, 413, 2031

\bibitem[{Metzger {et~al.}(1997)Metzger, Djorgovski, Kulkarni, Steidel, Adelberger, Frail, Costa, \& Frontera}]{Metzger1997}
Metzger, M.~R., Djorgovski, S.~G., Kulkarni, S.~R., {et~al.} 1997, \nat, 387, 878

\bibitem[{Minaev \& Pozanenko(2019)}]{Minaev_2019}
Minaev, P.~Y. \& Pozanenko, A.~S. 2019, \mnras, 492, 1919–1936

\bibitem[{Mooley {et~al.}(2018)Mooley, Deller, Gottlieb, Nakar, Hallinan, Bourke, Frail, Horesh, Corsi, \& Hotokezaka}]{Mooley_2018}
Mooley, K.~P., Deller, A.~T., Gottlieb, O., {et~al.} 2018, \nat, 561, 355

\bibitem[{{Musolino} {et~al.}(2024{\natexlab{a}}){Musolino}, {Duqu{\'e}}, \& {Rezzolla}}]{Musolino2024ApJ...966L..31M}
{Musolino}, C., {Duqu{\'e}}, R., \& {Rezzolla}, L. 2024{\natexlab{a}}, \apjl, 966, L31

\bibitem[{{Musolino} {et~al.}(2024{\natexlab{b}}){Musolino}, {Ecker}, \& {Rezzolla}}]{Musolino2024ApJ...962...61M}
{Musolino}, C., {Ecker}, C., \& {Rezzolla}, L. 2024{\natexlab{b}}, \apj, 962, 61

\bibitem[{Norris \& Bonnell(2006)}]{norris2006}
Norris, J.~P. \& Bonnell, J.~T. 2006, \apj, 643, 266

\bibitem[{Norris {et~al.}(2010)Norris, Gehrels, \& Scargle}]{norris2010}
Norris, J.~P., Gehrels, N., \& Scargle, J.~D. 2010, \apj, 717, 411

\bibitem[{{Nousek} {et~al.}(2006){Nousek}, {Kouveliotou}, {Grupe}, {Page}, {Granot}, {Ramirez-Ruiz}, {Patel}, {Burrows}, {Mangano}, {Barthelmy}, {Beardmore}, {Campana}, {Capalbi}, {Chincarini}, {Cusumano}, {Falcone}, {Gehrels}, {Giommi}, {Goad}, {Godet}, {Hurkett}, {Kennea}, {Moretti}, {O'Brien}, {Osborne}, {Romano}, {Tagliaferri}, \& {Wells}}]{Nousek2006ApJ...642..389N}
{Nousek}, J.~A., {Kouveliotou}, C., {Grupe}, D., {et~al.} 2006, \apj, 642, 389

\bibitem[{{O'Connor} {et~al.}(2024){O'Connor}, {Beniamini}, \& {Gill}}]{Ocon24}
{O'Connor}, B., {Beniamini}, P., \& {Gill}, R. 2024, arXiv e-prints, arXiv:2406.05297

\bibitem[{O'Connor {et~al.}(2022)O'Connor, Troja, Dichiara, Beniamini, Cenko, Kouveliotou, Gonz{\'{a} }lez, Durbak, Gatkine, Kutyrev, Sakamoto, S{\'{a}}nchez-Ram{\'{\i}}rez, \& Veilleux}]{oconnor2022}
O'Connor, B., Troja, E., Dichiara, S., {et~al.} 2022, \mnras, 515, 4890

\bibitem[{Oganesyan {et~al.}(2020)Oganesyan, Ascenzi, Branchesi, Salafia, Dall’Osso, \& Ghirlanda}]{Oganesyan_2020}
Oganesyan, G., Ascenzi, S., Branchesi, M., {et~al.} 2020, \apj, 893, 88

\bibitem[{Oppenheimer \& Volkoff(1939)}]{Oppenheimer_PhysRev.55.374}
Oppenheimer, J.~R. \& Volkoff, G.~M. 1939, Phys. Rev., 55, 374

\bibitem[{{Palenzuela} {et~al.}(2022){Palenzuela}, {Aguilera-Miret}, {Carrasco}, {Ciolfi}, {Kalinani}, {Kastaun}, {Mi{\~n}ano}, \& {Vigan{\`o}}}]{Palenzuela2022}
{Palenzuela}, C., {Aguilera-Miret}, R., {Carrasco}, F., {et~al.} 2022, \prd, 106, 023013

\bibitem[{{Perna} {et~al.}(2006){Perna}, {Armitage}, \& {Zhang}}]{Perna2006}
{Perna}, R., {Armitage}, P.~J., \& {Zhang}, B. 2006, \apjl, 636, L29

\bibitem[{{Piro} {et~al.}(2017){Piro}, {Giacomazzo}, \& {Perna}}]{Piro_2017}
{Piro}, A.~L., {Giacomazzo}, B., \& {Perna}, R. 2017, \apjl, 844, L19

\bibitem[{{Proga} \& {Zhang}(2006)}]{Proga2006}
{Proga}, D. \& {Zhang}, B. 2006, \mnras, 370, L61

\bibitem[{Punturo {et~al.}(2010)Punturo, Abernathy, Acernese, Allen, Andersson, Arun, Barone, Barr, Barsuglia, Beker, Beveridge, Birindelli, Bose, Bosi, Braccini, Bradaschia, Bulik, Calloni, Cella, Mottin, Chelkowski, Chincarini, Clark, Coccia, Colacino, Colas, Cumming, Cunningham, Cuoco, Danilishin, Danzmann, Luca, Salvo, Dent, Rosa, Fiore, Virgilio, Doets, Fafone, Falferi, Flaminio, Franc, Frasconi, Freise, Fulda, Gair, Gemme, Gennai, Giazotto, Glampedakis, Granata, Grote, Guidi, Hammond, Hannam, Harms, Heinert, Hendry, Heng, Hennes, Hild, Hough, Husa, Huttner, Jones, Khalili, Kokeyama, Kokkotas, Krishnan, Lorenzini, Lück, Majorana, Mandel, Mandic, Martin, Michel, Minenkov, Morgado, Mosca, Mours, Müller–Ebhardt, Murray, Nawrodt, Nelson, Oshaughnessy, Ott, Palomba, Paoli, Parguez, Pasqualetti, Passaquieti, Passuello, Pinard, Poggiani, Popolizio, Prato, Puppo, Rabeling, Rapagnani, Read, Regimbau, Rehbein, Reid, Rezzolla, Ricci, Richard, Rocchi, Rowan, Rüdiger, Sassolas, Sathyaprakash, Schnabel, Schwarz,
  Seidel, Sintes, Somiya, Speirits, Strain, Strigin, Sutton, Tarabrin, Thüring, van~den Brand, van Leewen, van Veggel, van~den Broeck, Vecchio, Veitch, Vetrano, Vicere, Vyatchanin, Willke, Woan, Wolfango, \& Yamamoto}]{Punturo_2010}
Punturo, M., Abernathy, M., Acernese, F., {et~al.} 2010, Classical and Quantum Gravity, 27, 194002

\bibitem[{{Reisenegger}(2009)}]{Reis09}
{Reisenegger}, A. 2009, \aap, 499, 557

\bibitem[{Rezzolla {et~al.}(2018)Rezzolla, Most, \& Weih}]{Rezzolla_2018}
Rezzolla, L., Most, E.~R., \& Weih, L.~R. 2018, \apjl, 852, L25

\bibitem[{{Rocha} {et~al.}(2023){Rocha}, {Horvath}, {de S{\'a}}, {Chinen}, {Bar{\~a}o}, \& {de Avellar}}]{Rocha23}
{Rocha}, L.~S., {Horvath}, J.~E., {de S{\'a}}, L.~M., {et~al.} 2023, Universe, 10, 3

\bibitem[{{Ronchini} {et~al.}(2023){Ronchini}, {Stratta}, {Rossi}, {Kann}, {Oganeysan}, {Dall'Osso}, {Branchesi}, \& {De Cesare}}]{Ronchini1}
{Ronchini}, S., {Stratta}, G., {Rossi}, A., {et~al.} 2023, \aap, 675, A117

\bibitem[{Rossi {et~al.}(2022)Rossi, Rothberg, Palazzi, Kann, D’Avanzo, Amati, Klose, Perego, Pian, Guidorzi, Pozanenko, Savaglio, Stratta, Agapito, Covino, Cusano, D’Elia, Pasquale, Valle, Kuhn, Izzo, Loffredo, Masetti, Melandri, Minaev, Guelbenzu, Paris, Paiano, Plantet, Rossi, Salvaterra, Schulze, Veillet, \& Volnova}]{Rossi_2022}
Rossi, A., Rothberg, B., Palazzi, E., {et~al.} 2022, \apj, 932, 1

\bibitem[{{Rossi} {et~al.}(2020){Rossi}, {Stratta}, {Maiorano}, {Spighi}, {Masetti}, {Palazzi}, {Gardini}, {Melandri}, {Nicastro}, {Pian}, {Branchesi}, {Dadina}, {Testa}, {Brocato}, {Benetti}, {Ciolfi}, {Covino}, {D'Elia}, {Grado}, {Izzo}, {Perego}, {Piranomonte}, {Salvaterra}, {Selsing}, {Tomasella}, {Yang}, {Vergani}, {Amati}, \& {Stephen}}]{rossi2020}
{Rossi}, A., {Stratta}, G., {Maiorano}, E., {et~al.} 2020, \mnras, 493, 3379

\bibitem[{{Rosswog}(2007)}]{Rosswog2007}
{Rosswog}, S. 2007, \mnras, 376, L48

\bibitem[{{Rouco Escorial} {et~al.}(2023){Rouco Escorial}, {Fong}, {Berger}, {Laskar}, {Margutti}, {Schroeder}, {Rastinejad}, {Cornish}, {Popp}, {Lally}, {Nugent}, {Paterson}, {Metzger}, {Chornock}, {Alexander}, {Cendes}, \& {Eftekhari}}]{RoucoEscorial_2023ApJ...959...13R}
{Rouco Escorial}, A., {Fong}, W., {Berger}, E., {et~al.} 2023, \apj, 959, 13

\bibitem[{{Rowlinson} {et~al.}(2013){Rowlinson}, {O'Brien}, {Metzger}, {Tanvir}, \& {Levan}}]{rowlinson2013}
{Rowlinson}, A., {O'Brien}, P.~T., {Metzger}, B.~D., {Tanvir}, N.~R., \& {Levan}, A.~J. 2013, \mnras, 430, 1061

\bibitem[{{Sari} {et~al.}(1998){Sari}, {Piran}, \& {Narayan}}]{sari1998}
{Sari}, R., {Piran}, T., \& {Narayan}, R. 1998, \apjl, 497, L17

\bibitem[{{Sarin} {et~al.}(2019){Sarin}, {Lasky}, \& {Ashton}}]{Sarin2}
{Sarin}, N., {Lasky}, P.~D., \& {Ashton}, G. 2019, \apj, 872, 114

\bibitem[{{Sarin} {et~al.}(2020){Sarin}, {Lasky}, \& {Ashton}}]{Sarin1}
{Sarin}, N., {Lasky}, P.~D., \& {Ashton}, G. 2020, \prd, 101, 063021

\bibitem[{Shibata \& Hotokezaka(2019)}]{shibata2019}
Shibata, M. \& Hotokezaka, K. 2019, Annual Review of Nuclear and Particle Science, 69, 41

\bibitem[{Stratta {et~al.}(2018{\natexlab{a}})Stratta, Ciolfi, Amati, Bozzo, Ghirlanda, Maiorano, Nicastro, Rossi, Vinciguerra, Frontera, Götz, Guidorzi, O’Brien, Osborne, Tanvir, Branchesi, Brocato, Dainotti, De~Pasquale, Grado, Greiner, Longo, Maio, Mereghetti, Mignani, Piranomonte, Rezzolla, Salvaterra, Starling, Willingale, Böer, Bulgarelli, Caruana, Colafrancesco, Colpi, Covino, D’Avanzo, D’Elia, Drago, Fuschino, Gendre, Hudec, Jonker, Labanti, Malesani, Mundell, Palazzi, Patricelli, Razzano, Campana, Rosati, Rodic, Szécsi, Stamerra, van Putten, Vergani, Zhang, \& Bernardini}]{Stratta_2018}
Stratta, G., Ciolfi, R., Amati, L., {et~al.} 2018{\natexlab{a}}, Advances in Space Research, 62, 662–682

\bibitem[{Stratta {et~al.}(2018{\natexlab{b}})Stratta, Dainotti, Dall’Osso, Hernandez, \& Cesare}]{stratta2018}
Stratta, G., Dainotti, M.~G., Dall’Osso, S., Hernandez, X., \& Cesare, G.~D. 2018{\natexlab{b}}, \apj, 869, 155

\bibitem[{Tang {et~al.}(2019)Tang, Huang, Geng, \& Zhang}]{Tang_2019}
Tang, C.-H., Huang, Y.-F., Geng, J.-J., \& Zhang, Z.-B. 2019, \apjs, 245, 1

\bibitem[{{Tsvetkova} {et~al.}(2017){Tsvetkova}, {Frederiks}, {Golenetskii}, {Lysenko}, {Oleynik}, {Pal'shin}, {Svinkin}, {Ulanov}, {Cline}, {Hurley}, \& {Aptekar}}]{tsvetkova_2017}
{Tsvetkova}, A., {Frederiks}, D., {Golenetskii}, S., {et~al.} 2017, \apj, 850, 161

\bibitem[{{Tsvetkova} {et~al.}(2021){Tsvetkova}, {Frederiks}, {Svinkin}, {Aptekar}, {Cline}, {Golenetskii}, {Hurley}, {Lysenko}, {Ridnaia}, \& {Ulanov}}]{tsvetkova_2021}
{Tsvetkova}, A., {Frederiks}, D., {Svinkin}, D., {et~al.} 2021, \apj, 908, 83

\bibitem[{{Urrutia} {et~al.}(2021){Urrutia}, {De Colle}, {Murguia-Berthier}, \& {Ramirez-Ruiz}}]{Urru21}
{Urrutia}, G., {De Colle}, F., {Murguia-Berthier}, A., \& {Ramirez-Ruiz}, E. 2021, \mnras, 503, 4363

\bibitem[{{Usov}(1992)}]{usov1992}
{Usov}, V.~V. 1992, \nat, 357, 472

\bibitem[{Yi {et~al.}(2016)Yi, Xi, Yu, Wang, Mu, Lü, \& Liang}]{Yi_2016}
Yi, S.-X., Xi, S.-Q., Yu, H., {et~al.} 2016, \apjs, 224, 20

\bibitem[{{Zhang} {et~al.}(2015){Zhang}, van {Eerten}, Burrows, Ryan, Evans, Racusin, Troja, \& MacFadyen}]{Zhang_2015}
{Zhang}, van {Eerten}, H., Burrows, D.~N., {et~al.} 2015, \apj, 806, 15

\bibitem[{{Zhang} {et~al.}(2006){Zhang}, {Fan}, {Dyks}, {Kobayashi}, {M{\'e}sz{\'a}ros}, {Burrows}, {Nousek}, \& {Gehrels}}]{ZhangB_2006ApJ...642..354Z}
{Zhang}, B., {Fan}, Y.~Z., {Dyks}, J., {et~al.} 2006, \apj, 642, 354

\bibitem[{{Zhang} {et~al.}(2008){Zhang}, {Woosley}, \& {Heger}}]{ZhangW2008ApJ...679..639Z}
{Zhang}, W., {Woosley}, S.~E., \& {Heger}, A. 2008, \apj, 679, 639

\bibitem[{{Zhu} {et~al.}(2023){Zhu}, {Liu}, {Shi}, {Ding}, {Sun}, \& {Zhang}}]{zhu_2023}
{Zhu}, S.-Y., {Liu}, Z.-Y., {Shi}, Y.-R., {et~al.} 2023, \apj, 950, 30

\bibitem[{Özel \& Freire(2016)}]{ozel_2016}
Özel, F. \& Freire, P. 2016, \araa, 54, 401–440

\end{thebibliography}

\clearpage
\onecolumn

\begin{appendix}

\section{Results of the light curve fit}
Table~\ref{tab:fit} lists the redshifts and best-fit parameters of the simple power-law (PL, 25 cases)/broken power-law (BPL, 15 cases) models for the 40 bursts in our LC fit sample, described in Sect.~\ref{sub:classification}.
\begin{table*}[h]
  \centering 
\caption{Best-fit parameters of the simple (PL) and broken power-law (BPL) models for all the 40 SGRBs light curves in the LC fit sample (see Sect.~\ref{sub:classification}).}
\begin{tabular}{lccccccccc}
\hline
\hline
    GRB name & $z$   & $t_0$   & $\alpha_1$          & $\alpha_2$      & $t_{\mathrm{break}}$    & $F_{\mathrm{pl,norm}/\mathrm{break}}$  & $\chi^2$& $\nu$ &  $p$-value \\
             &     &  (s)   &                      &                 &  ($\times 10^3$ s)                    &($\times10^{-12}$ erg cm$^{-2}$ s$^{-1}$)&        &         \\
\hline
    050724(F)& 0.254  & 370  & $0.93\pm0.06$  &        -       &     -       & $(1.63\pm0.19)$              & 23.9 & 10   &     1.0\\
    061006   & 0.461  & 168  & $0.78\pm0.05$  &        -       &     -       & $(1.45\pm0.13)$                                     & 13.3 & 9    &     1.0\\
    070724A  & 0.457  & 385  & $1.16\pm0.10$  &        -       &     -       & $(1.3\pm0.3)$  & 21.8 & 4    &$8.58\times 10^{-2}$\\ 
    070809   & 0.2187 & 126  & $0.50\pm0.06$  &        -       &     -       & $(1.7\pm0.2)$  & 43.0 & 14   &$1.14\times 10^{-3}$\\
    090426   & 2.609  & 120  & $0.95\pm0.03$  &        -       &     -       & $(2.85\pm0.19)$& 40.9 & 25   &$7.57\times 10^{-3}$  \\
    111117A  & 2.211  & 200  & $1.22\pm0.07$  &        -       &     -       & $(5.4\pm0.6)\times 10^{-1}$  & 4.8  & 5    &$7.99\times 10^{-1}$\\
    120804A  & 1.05   & 150  & $1.08\pm0.03$  &        -       &     -       & $(1.26\pm0.07)\times 10$& 55.1 & 30   &      1.0\\
    121226A  & 1.37   & 146  & $0.97\pm0.05$  &        -       &     -       & $(5.0\pm0.5)$  & 21.5 & 11   &$1.36\times 10^{-1}$\\
    131004A(F)& 0.717 & 105  & $1.00\pm0.06$  &        -       &     -       & $(4.9\pm0.7)$  & 25.5 & 8    &$1.60\times 10^{-1}$\\
    140129B  & 0.43   & 400  & $1.29\pm0.09$  &        -       &     -       & $(5.9\pm0.8)$  & 24.0 & 15   &$1.52\times 10^{-2}$\\
    140930B  & 1.465  & 214  & $1.75\pm0.10$  &        -       &     -       & $(1.4\pm0.2)$  & 61.7 & 23   &     1.0\\
    150423A  & 1.394  & 100  & $0.91\pm0.06$  &        -       &     -       & $(1.11\pm0.13)$& 4.5  & 6    &$6.47\times 10^{-2}$\\
    150831A  & 1.18   & 200  & $1.07\pm0.08$  &        -       &     -       & $(0.71\pm0.10)$& 2.2  & 4    &      1.0\\
    160303A  & 1.01   & 600  & $0.68\pm0.10$  &        -       &     -       & $(0.76\pm0.13)$& 15.2 & 5    &$2.90\times 10^{-1}$\\
    160525B  & 0.64   & 99   & $1.35\pm0.10$  &        -       &     -       & $(0.67\pm0.14)$& 27.4 & 11   &      1.0\\
    160821B  & 0.1619 & 300  & $1.33\pm0.13$  &        -       &     -       & $(2.0\pm0.6)$  & 16.2 & 3    & $7.94\times 10^{-1}$\\
    170728A  & 1.493  & 250  & $0.98\pm0.07$  &        -       &     -       & $(3.6\pm0.4)$  & 5.9  & 4    & $4.76\times 10^{-1}$\\
    180418A  & 1.56   & 3170 & $0.84\pm0.04$  &        -       &     -       & $(2.2\pm0.15)$                 & 17.8 & 16   &       1.0\\
    180727A  & 1.95   & 100  & $1.20\pm0.09$  &        -       &     -       & $(0.9\pm0.2)$  & 6.1  & 3    &       1.0\\
    180805B  & 0.6612 & 479  & $1.16\pm0.11$  &        -       &     -       & $(1.3\pm0.2) $ & 16.3  & 8   &       1.0  \\
    191019A  & 0.248  & 3545 & $1.13\pm0.13$  &        -       &     -       & $(1.22\pm0.13)$                & 9.1  & 6    &$8.19\times 10^{-2}$\\
    200411A  & 0.82   & 400  & $0.84\pm0.08$  &        -       &     -       & $(2.0\pm0.2)$  & 18.7 & 9    &       1.0\\
    200522A  & 0.5536 & 450  & $0.64\pm0.08$  &        -       &     -       & $(1.3\pm0.2)$  & 6.6  & 3    &$7.40\times 10^{-1}$\\   
    210726A  & 0.37   & 500  & $0.56\pm0.02$  &        -       &     -       & $(1.41\pm0.08)$& 2.5  & 7    &$9.90\times 10^{-1}$\\
    211023B & 0.862  & 750  & $0.78\pm0.06$   &        -       &     -       & $(1.48\pm0.15)$   & 8.9  & 6    &       1.0\\
\hline
    051221A  & 0.5464 & 314  & $0.65\pm0.03$  & $1.44\pm0.08$   & $6.16\times10$  & $(5.5\pm0.4)\times 10^{-1}$& 96.6  & 60  & $1.26\times 10^{-8}$ \\
    060614   & 0.125  & 4000 & $0.06\pm0.03$  & $1.81\pm0.03$   & $4.45\times10$  & $(5.8\pm0.2)$& 172.9& 149 & $1.11\times 10^{-16}$ \\
    061201   & 0.111  & 80   & $0.65\pm0.06$  & $2.12\pm0.10$   & $3.00$  & $(3.9\pm0.5)\times 10$& 29.3 & 24  & $2.50\times 10^{-9}$ \\
    070714B  & 0.923  & 300  & $0.65\pm0.12$  & $2.11\pm0.10$   & $2.13$  & $(1.8\pm0.2)\times 10$& 44.8 & 25  & $2.43\times 10^{-6}$ \\
    090510   & 0.903  & 100  & $0.66\pm0.03$  & $2.28\pm0.06$   & $1.67$  & $(7.6\pm0.4)\times 10$& 93.4 & 99  & $1.11\times 10^{-16}$ \\
    110402A  & 0.854  & 593  & $0.48\pm0.06$  & $2.25\pm0.15$   & $8.43$  & $(2.0\pm0.2)$& 13.7 & 15 & $6.11\times 10^{-6}$ \\
    130603B  & 0.3568 & 70   & $0.38\pm0.03$  & $1.69\pm0.06$   & $2.90$  & $(3.3\pm0.2)\times 10$& 128.8& 69  & $1.11\times 10^{-16}$\\
    140903A  & 0.3529 & 200  & $0.15\pm0.03$  & $1.25\pm0.05$   & $9.65$  & $(9.7\pm0.5)$& 28.2 & 34  & $8.55\times 10^{-15}$ \\
    150424A  & 0.3    & 453  & $0.76\pm0.02$  & $2.4\pm0.3$     & $2.03\times10^2$  & $(3.9\pm0.3)\times 10^{-1}$& 23.3 & 32  & $1.24\times 10^{-6}$ \\    
    151229A  & 0.63   & 90   & $0.26\pm0.19$  & $0.96\pm0.04$   & $3.47\times10^{-1}$  & $(2.8\pm0.9)\times 10^{2}$& 73.1 & 70  & $9.51\times 10^{-7}$\\
    161001A  & 0.67   & 207  & $0.75\pm0.05$  & $1.37\pm0.05$   & $3.53$  & $(4.9\pm0.4)\times 10$& 57.3 & 45  & $1.45\times 10^{-5}$\\
    170728B  & 1.272  & 400  & $0.53\pm0.03$  & $1.34\pm0.02$   & $2.53$  & $(2.1\pm0.1)\times 10^2$& 213.6& 193 & $1.11\times 10^{-16}$ \\
    180618A  & 0.52   & 80   & $0.11\pm0.40$  & $1.77\pm0.04$   & $1.27\times10^{-1}$  & $(1.8\pm0.2)$              & 130.4& 97  & $2.98\times 10^{-11}$ \\
    210323A  & 0.733  & 800  & $0.50\pm0.06$  & $3.4\pm0.3$     & $1.32\times10$  & $(4.2\pm0.4)$& 10.4 & 10  & $1.22\times 10^{-5}$ \\
    211211A  & 0.0763 & 3400 & $-0.12\pm0.18$ & $2.07\pm0.08$   & $7.97$  & $(5.5\pm0.6)\times 10$& 72.9 & 47  & $2.39 \times 10^{-10}$ \\
 \hline
\label{tab:fit}
\end{tabular}
\tablefoot{The first 25 bursts 
are the ones for which a power-law model was enough to describe the light curve behaviour (PL subsample). 
The last 15 bursts are the BPL subsample, for which the addition of a break resulted statistically significant (see Sec.~\ref{sub:fit}). 
The $(\mathrm{F})$ marks the cases for which a flaring component was discarded to allow a better fit of the afterglow.}

\end{table*}

\newpage
\section{SGRB sample burst energetics}
Table~\ref{tab:eiso} quotes the isotropic equivalent energy $E_{\mathrm{iso}}$ for 42/85 SGRBs in the initial sample for which 
this value is publicly available in the literature. Depending on the reference, the range of energies chosen to compute $E_{\rm iso}$ is not the same. Bold entries are the 15 SGRBs included in the plateau subsample. In Fig.~\ref{fig:Eiso}, these values are presented graphically, with particular emphasis on the cases where a plateau was identified, which are highlighted in azure.

\begin{table}[H]
        \begin{minipage}{0.5\linewidth}
                \caption{Values of $E_{\mathrm{iso}}$ for 42 SGRBs of the initial sample which are publicly available in the literature.}
                \label{table:student}
                \centering
                \begin{tabular}{lcc}
\hline
\hline
GRB name & $E_{\mathrm{iso}}$ & Reference \\
    &($10^{52}$ erg)&\\
     \hline
     050724      & $0.024$   & 1 \\
            \textbf{051221A}     & $0.31$   & 2 \\
            060313      & $2.9$   & 3 \\
            \textbf{060614}      & $0.27$   & 2 \\
            060801      & $0.478$   & 2 \\
            061006      & $0.21$   & 2 \\
            \textbf{061201}      & $0.017$   & 2 \\
            \textbf{070714B}     & $0.64$   & 2 \\
            070724A     & $0.03$   & 3 \\
            070809      & $0.09$   & 3 \\
            071227      & $0.059$   & 2 \\
            080905A     & $0.02$   & 3 \\
            081226A     & $0.09$   & 3 \\
            090426      & $0.24$   & 1 \\
            \textbf{090510}      & $5.71$   & 2 \\
            091109B     & $0.18$   & 3 \\
            100117A     & $0.22$   & 3 \\
            100206A     & $0.051$   & 2 \\
            101219A     & $0.651$   & 2 \\
            110112A     & $0.03$   & 3 \\
            \textbf{110402A}     & $1.52$   & 4 \\
            111117A     & $0.55$   & 3 \\
            120804A     & $0.657$   & 2 \\
            121226A     & $0.37$   & 3 \\
            \textbf{130603B}     & $0.196$   & 2 \\
            130912A     & $0.16$   & 3 \\
            131004A     & $0.138$   & 1 \\
            140129B     & $0.07$   & 3 \\
            140516A     & $0.02$   & 3 \\
            140622A     & $0.07$   & 3 \\
            \textbf{140903A}     & $0.08$   & 3 \\
            140930B     & $0.40$   & 3 \\
            150101B     & $0.004$   & 3 \\
            \textbf{150424A}     & $0.434$   & 2 \\
            \textbf{151229A}     & $0.12$   & 5 \\
            160410A     & $9.3$   & 2 \\
            \textbf{161001A}     & $0.30$   & 5 \\
            \textbf{170728B}     & $0.40$   & 5 \\
            \textbf{180618A}     & $0.39$   & 5 \\
            \textbf{210323A}     & $0.43$   & 5 \\
            \textbf{211211A}     & $1.24$   & 5 \\
            191019A     & $0.1$   & 6\\
\hline
\label{tab:eiso}
\end{tabular}
\tablefoot{Bold entries are the bursts belonging to the plateau subsample.}
\tablebib{
   (1) \citet{tsvetkova_2021}; (2) \citet{tsvetkova_2017}; (3) \citet{fong_2015}; (4) \citet{Minaev_2019};
   (5) \citet{zhu_2023}; (6) \citet{Lazzati_2023}.
   }
        \end{minipage}\hfill
        \begin{minipage}{0.45\linewidth}
                \centering
                \includegraphics[width=1.0\columnwidth,angle=0]{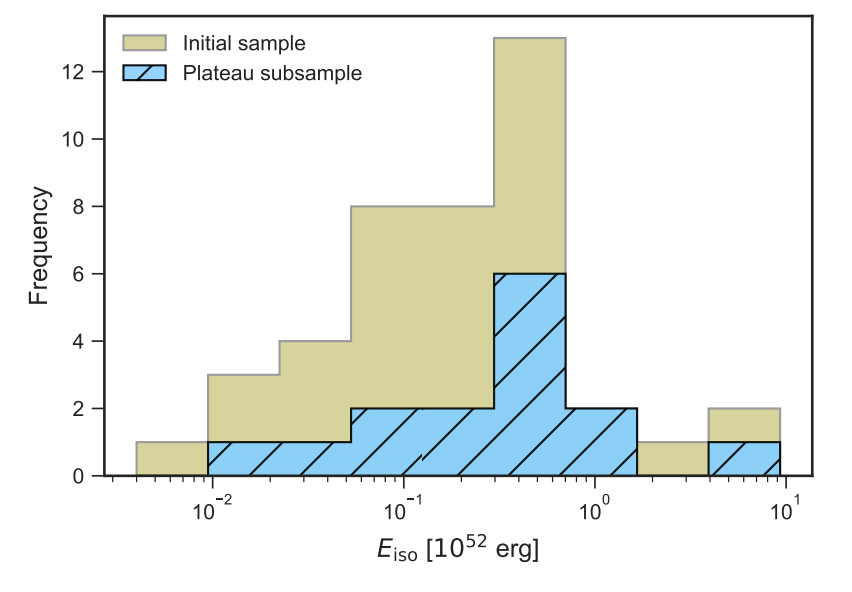}
                \captionof{figure}{Isotropic equivalent energy distribution ($E_{\mathrm{iso}}$) of the SGRBs in the plateau subsample compared with that of the whole sample (see Sect.~\ref{sub:bias} and Table~\ref{tab:eiso}). }
                \label{fig:Eiso}
        \end{minipage}
\end{table}

\newpage

\onecolumn

\section{Assumptions in the magnetar model}
\label{app:magnetarmodel}
 In the magnetar model formalism by \citet{dallosso2011}, the radiative efficiency $k' = 4\epsilon_e (d\ln t/d \ln T)$ (see Eq. 7) encodes our ignorance of the microphysics and of the surrounding environment (i.e. wind or constant interstellar matter). In this work, $k'$ is generically kept $<1$, where $\epsilon_e$ is the fraction of the total energy released in the merger that is transferred to the radiating electrons, while the factor $(d\ln t/d \ln T)$ encloses the hydrodynamical evolution of the shock: the condition on $k'$ implies $\epsilon_e <0.5$. The extension of the formalism that includes resistive effects in the magnetosphere \citep{stratta2018}, introduces a new parameter $\alpha$ that governs the spin-down luminosity. In this work, we fixed $\alpha$ at 0. Ultimately, the total moment of inertia of the magnetar is taken as the approximation in \citet{LATTIMER_2007}. We adopted the standard values for the NS mass ($M_{\mathrm{NS}}= 1.4\,$M$_{\odot}$) and radius ($R_{\mathrm{NS}}= 12 $ km).

\section{Minimum prompt luminosity and $\kappa$-criterion}
\label{app:Lmin}
The $\kappa$-criterion was applied to the EE-rejected SGRBs building upon the work of \cite{Dall_Osso_2023}, 
to infer the presence or absence of a plateau feature. 
Here, $\kappa$ is defined as the ratio between the intrinsic minimum luminosity in the prompt $L_{\gamma,\mathrm{min}}$ and the isotropic equivalent luminosity of the plateau $L_{\mathrm{sd}}$ (see Sect.~\ref{subsec:criteria}):

\begin{equation}
        \kappa=L_{\gamma,\mathrm{min}}/L_{\mathrm{sd}}\approx1.2\times 10^5 \,\epsilon \,P^{5/3} (R_6\, M_{1.4}^{2/3})^{-1} ,
        \label{eq:ratio1}
\end{equation}
with $\epsilon$ the radiative efficiency (typically $\sim 0.1$) of the prompt emission, $P$ the NS spin period in seconds, $R_6= R_{\mathrm{NS}}/(10^6 \,\mathrm{cm})$, $M_{1.4}= M_{\mathrm{NS}}/(1.4 \,$M$_{\odot})$. For given NS mass and radius, $\kappa$ is a function of the NS spin period alone; thus values of $\kappa>30$ are considered unphysical. To derive $L_{\gamma,\mathrm{min}}$ and $L_{\mathrm{sd}}$, we analysed the rest-frame luminosity light curves of all the 19 SGRBs in the EE-~rejected sample. $L_{\gamma,\mathrm{min}}$ was determined by identifying the luminosity of the final data point of the prompt emission phase preceding the steep decay, as described by criterion (3) in Sect.~\ref{sub:classification}, and then dividing it by the jet beaming factor to obtain the intrinsic luminosity. We assumed an average beaming factor of $f_b=0.01$, which is reasonable for SGRBs and corresponds to a $\theta_j\sim 8$ deg. For $L_{\mathrm{sd}}$, we assumed it coincides with the minimum detected light curve luminosity $L_{p,\mathrm{min}}$, since $L_{p,\mathrm{min}}\gtrsim L_{\mathrm{sd}}$ by definition. This approach allowed us to calculate $\kappa$ for each burst in the EE-rejected subsample and we concluded that in 9 of these cases, the presence of a magnetar could be confidently ruled out.

\end{appendix}
\end{document}